\documentclass[twocolumn,prb]{revtex4}
\usepackage{epsfig}
\usepackage{epstopdf}
\usepackage{amsmath}
\usepackage[latin1]{inputenc}
\usepackage{graphicx}
\usepackage{color}

\allowdisplaybreaks[1]

\newcommand{\nc}{\newcommand}
 \nc{\tcb}{\textcolor{blue}}  
 \nc{\tcr}{\textcolor{red}}
 \nc{\be}{\begin{equation}} 
 \nc{\ee}{\end{equation}}
 \nc{\bt}{\begin{tabular}} 
 \nc{\et}{\end{tabular}}
 \nc{\bea}{\begin{eqnarray}}  
 \nc{\eea}{\end{eqnarray}}
 \nc{\ba}{\begin{array}}  
 \nc{\ea}{\end{array}}
 \nc{\rds}{{\rm d}s} 
 \nc{\rdt}{{\rm d}t} 
 \nc{\rdr}{{\rm d}r}
 \nc{\rdO}{{\rm d}\Omega} 
 \nc{\s}{{\rm S}} 
 \nc{\dis}{\displaystyle} 
 \nc{\crit}{_{\rm cr}} 
 \nc{\rd}{{\rm d}}
 \nc{\munu}{{\mu\nu}} 
 \nc{\erm}{{\rm e}}
 \nc{\drm}{{\rm d}}
 \nc{\ov}{\overline}

\begin{document}

\title{The physical leaky tank car problem, revisited}%

\author{S. Esposito}
\email{Salvatore.Esposito@na.infn.it}%
\affiliation{
Istituto Nazionale di Fisica Nucleare, Sezione di Napoli, Complesso Universitario di Monte
S.\,Angelo, via Cinthia, I-80126 Naples, Italy}
\author{Matteo Olimpo}%
\email{olimpomatteo@gmail.com}%
\affiliation{
Via Anfiteatro 3, I-80078 Pozzuoli, Italy}

%\thanks{}%
%\subjclass{}%
%\keywords{}%

%\date{}%
%\dedicatory{}%
%\commby{}%
%----------------------------------------------------------------
\begin{abstract}
\noindent An exhaustive analysis of the leaky tank car problem is presented, pointing out its intriguing physical properties, which well serve to students and teachers for illustrating standard Newtonian mechanics in a highly non-standard fashion. Calculus (at a leading undergraduate level) is effectively required only to examine some details concerning the solution of the equation of motion and interesting limiting cases. Instead we let any student to appreciate all the physical content of the problem, within a proper simple model and its generalization, ranging from the motion of the leaky tank car to that of the water flowing from the car with their peculiarities. Generalizations of the problem to more than one draining hole (even with different sizes), as well as to ``two-dimensional" geometries, reveal further intriguing results, culminating into a no-rotation theorem and its corollaries, thus rendering unique the problem at hand for allowing students to fully recognize the power of physical analysis.
\end{abstract}

\maketitle

%----------------------------------------------------------------

\section{Introduction}

\noindent Quite recently, a certain interest has been revived \cite{ekman} about a not so old problem, known as the leaky tank car problem \cite{mcdonald}, aimed at describing the motion of a tank car filled with a liquid that is draining out (vertically relative to the car) of an off-center hole.\cite{mcdonald2} The interest is, likely, only didactic in nature, since practical observations of the phenomenon are difficult and, seemingly, only relegated to a physics teaching lab. Nevertheless, the theoretical exercise behind the problem is quite interesting for Newton dynamics, and is usually assumed to be surprisingly complex. The correct and complete quantitative description of the motion\cite{mcdonald,mcdonald2} (under general assumptions concerning the fluid flow) is, indeed, perceived as mathematically involved, and this allowed the literature\cite{ekman,mcdonald2} to be enriched by toy models mainly aimed at illuminating the physical content of an interesting problem.  

Convinced by the genuine interest of the problem at hand, %which can be certainly considered as an important resource for teaching basic principles of classical mechanics (including energy and momentum conservation laws, etc.) and even to illustrate to students the usual way of reasoning of physicists dealing with a given problem, 
we have approached it by a different point of view, centered around the physical aspects of the problem, rather than the (more reassuring) mathematical ones. In such a way, not only we will recover known results in a more physically transparent manner, but we will be able as well to get new insight into the problem, by exploring intriguing generalizations and unveiling a number of features that even render far more interesting the problem itself, especially for teaching purposes. Also, we will show that the mathematics required can be essentially limited to a minimum, that is -- with the exception of finding the solution of a differential equation (of course the same found by McDonald\cite{mcdonald}) -- even to a pre-calculus stage. A non-negligible part of the results obtained will come from fair ``numerical" analysis (a high-sounding term to denote just graphical plotting of given well-defined mathematical quantities), which well exemplifies standard work of (today) physicists aimed at extracting physical information from mathematical results. Thus, contrary to what believed earlier, the leaky tank car problem will reveal to be not at all a complex problem, but it is well suitable for teaching marvelous, non trivial physics to undergraduate students. 

Where applicable, in the following we will always tend to use a pre-calculus formalism, although the careful reader will be able to easily translate it into a more formal calculus one. 

\section{Physical insight into the McDonald model}

\noindent Let us consider a tank car of mass $m$ and horizontal surface area $S_c$, with a small hole of surface area $S_h$ (with $S_h \ll S_c$) and located horizontally at distance $D$ from the center of the tank (see Fig.$\,$\ref{fig1}). Initially, at time $t=0$ the tank car is filled with water (or any other liquid) of mass $M_0$, while we denote with $M(t)$ the mass of the water inside the car at time $t$. $x(t)$  is instead the horizontal coordinate of the center of the car, which is assumed to start to move from the rest at $x=0$ at time $t=0$. 

\subsection{Water draining}

\begin{figure}[t]
\includegraphics[width=7cm]{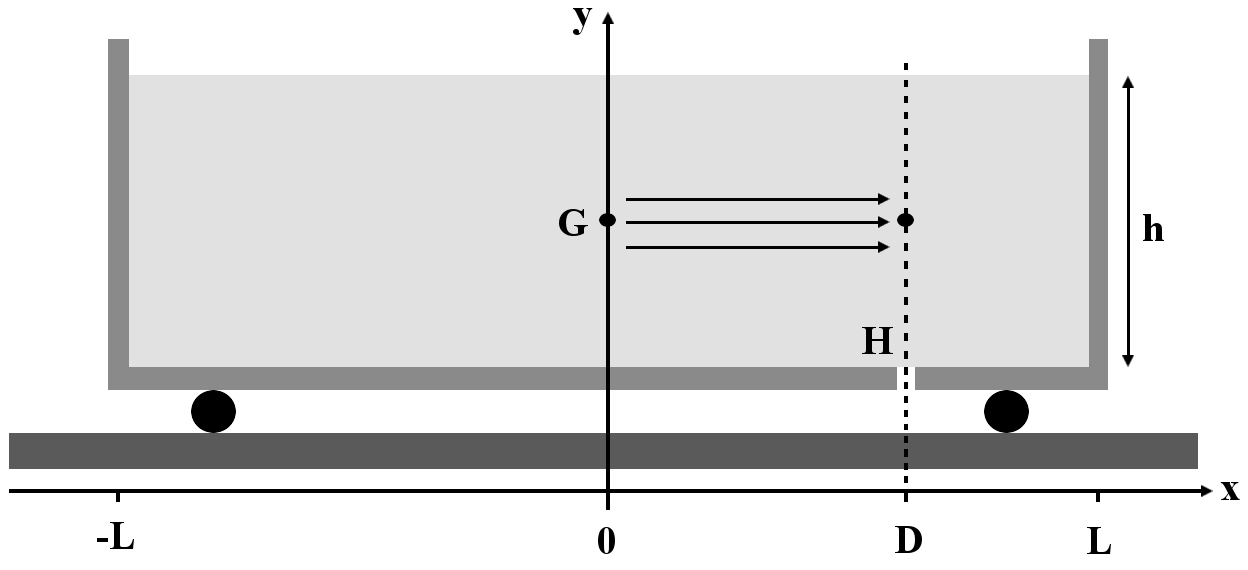}
\caption{A leaky tank car (filled with liquid at height $h$) of length $2L$ with a draining small hole $H$ at position $D$ off center: water flows overall from its center of mass $G$ to the hole.}
\label{fig1}
\end{figure}

\noindent Three basic assumptions will be made:
\begin{itemize}
\item[A1] the water exits the drain with zero horizontal velocity relative to that of the tank car;
\item[A2] water is incompressible and inviscid;
\item[A3] the Bernoulli equation holds for the water, that is energy is conserved in the water flow and kinetic energy of the water in the tank (in the rest frame of the tank) can be neglected.
\end{itemize}
Assumption A2 implies that the continuity equation holds, so that the vertical velocity of water exiting the drain is given by ($\hat{v}_E = - \hat{z}$ in the reference frame in Fig.$\,$\ref{fig1}):
\be
v_E \, = \, - \frac{S_c}{S_h} \, \dot{h} \, ,
\label{ce}
\ee
where $\dot{h} = \dot{h}(t)$ is the velocity with which the free surface of the water lowers with time.\footnote{Here and below, a dot above a given quantity denotes differentiation with respect to time, or rather a finite variation of that quantity with respect to time if the problem is illustrated without making recourse to calculus.} Instead, from assumption A3 follows that the Torricelli law holds:
\be
v_E \, = \, \sqrt{2 g h} \, .
\label{tl}
\ee
From these two equations, the variation of water height $h(t)$ with time may be easily deduced even without employing calculus. Indeed, while assuming finite variations, by equating (\ref{ce}) and (\ref{tl}) we obtain $2 \Delta \sqrt{h} = - \frac{S_h}{S_c} \sqrt{2 g} \, \Delta t$, so that $2 \sqrt{h} = - \frac{S_h}{S_c} \sqrt{2 g} \, t + {\rm const.}$, where the additive constant may be deduced from the initial condition on the height (that is, $h=h_0$ for $t=0$, and thus const.$\, = 2 \sqrt{h_0}$). %: 
%\be
%\sqrt{h} \, = \, \sqrt{h_0} \left( 1 - \frac{S_h}{S_c} \, \sqrt{\frac{g}{2 h_0}} \, t \right) .
%\ee
It is quite instructive to introduce the {\it emptying time} $t_s$ defined by $h(t_s) = 0$, that is
\be
t_s \, = \, \frac{S_c}{S_h} \, \sqrt{\frac{2 h_0}{g}} \, ;
\label{ts}
\ee
in such a way, the time equation $h=h(t)$ for the water height can be written as:
\be
h \, = \, h_0 \left( 1 - \frac{t}{t_s} \right)^2 \!\!\! .
\label{ht}
\ee
The time equation for the mass $M=M(t)$ of the water inside the car immediately follows by introducing the water density $\rho$, i.e. $M = \rho S_c h$ (and thus $M_0 = \rho S_c h_0$):
\be
M \, = \, M_0 \left( 1 - \frac{t}{t_s} \right)^2 \!\!\! .
\label{mt}
\ee
Interesting as well is the mass rate $\dot{M} = \rho S_c \dot{h} = - \rho S_h \sqrt{2 g h}$:
\be
\dot{M} \, = \, - \, \frac{2M_0}{t_s} \left( 1 - \frac{t}{t_s} \right) ,
\label{mpt}
\ee
whose initial value $\dot{M}_0 \equiv \dot{M}(0) = - 2 M_0/t_s$ is obviously negative (water drains from the hole), and, for future reference, the time variation of the mass rate $\ddot{M}$, resulting to be\footnote{From Eq.$\,$(\ref{mpt}): $\Delta \dot{M} = - \, \frac{2M_0}{t_s} \left( - \frac{\Delta t}{t_s} \right)$.} constant with time and positive:
\be
\ddot{M} \, = \, \frac{2M_0}{t_s^2} \, .
\label{mp2t}
\ee

\subsection{Equation of motion}

\noindent The dynamics of the system is described by the Newton law:
\be
M_{\rm tot} \, a \, = \, F ,
\label{nl}
\ee
where $F$ is the horizontal force acting on the system tank car + water, $a = \ddot{x}(t)$ is the acceleration of such system, and the total mass is given by $M_{\rm tot} =  m + M(t)$. The active force driving the system is, of course, gravity, which drains the water from the hole. According to McDonald,\cite{mcdonald} water moves horizontally inside the car flowing overall in the $+x$ direction (see Fig.$\,$\ref{fig1}) from the center of mass $G$ of the water (located at $x=0$) to the exiting hole $H$ (located at $x=D$).\footnote{Here and below we refer only to the horizontal component of this flow, since the vertical one does not affect the motion of the car.} Such a motion is due to the force $F$ developed as a constraint reaction by the car (inside) walls, acting overall in the $+x$ direction. As a consequence of the action-reaction principle, the force $F_w$ done by water upon the car is opposite to $F$, i.e. $F_w = - F$. Now, from assumption A3 it follows that water moves inside the tank car according to a stationary regime since, otherwise, the Bernoulli equation should not hold. As a result, within the assumptions made, the force $F_w$ has to be {\it constant} (it does not depend on time), and so does $F$, so that the problem at hand is just a dynamic problem of a constant force acting on a variable mass system. This result, which in the previous literature\cite{mcdonald,mcdonald2} is not so transparent, is instead obvious when remembering that the active force (gravity) driving the system is constant in time.

Once established this, the expression for the force $F$ entering into Eq.$\,$(\ref{nl}) may be deduce as follows from the impulse-momentum theorem. The momentum $p_{w_{\rm ins}}$ of the water acting on the car walls overall in the $-x$ direction is given by the product of the mass element $\Delta M$ flowing in the time interval $\Delta t$ with velocity $v_w$, which is simply given in the model above by $D/\Delta t$:
\be
p_{w_{\rm ins}} = - \,  \Delta M \cdot v_w = - \, \Delta M \cdot \frac{D}{\Delta t} = - \, \dot{M}  D.
\label{pw}
\ee
The variation of such momentum in time equals the water impulse, $F_w \Delta t = \Delta p_{w_{\rm ins}}$, so that
\be
- \, F \cdot \Delta t = - \, D \, \Delta \dot{M} \, .
\ee
Since the force $F$ is constant from the beginning ($t=0$) to the end ($t=t_s$) of the motion, we have:
\be
F \cdot t_s = D \left( \dot{M}(t_s) - \dot{M}(0) \right) = - \, D \, \dot{M}_0 \, ,
\ee
from which the expression for the force immediately follows:
\be
F = - \, \frac{\dot{M}_0 D}{t_s}  = \frac{2 M_0 D}{t_s^2} = \ddot{M} D 
\label{force}
\ee
(where we have also used Eqs.$\,$(\ref{mpt}),(\ref{mp2t})).

By substituting Eq.$\,$(\ref{force}) into Eq.$\,$(\ref{nl}), the equation of motion of the system results to be:
\be
\left[ m + M(t) \right] \ddot{x}(t) = \frac{2 M_0 D}{t_s^2} \, .
\label{eom}
\ee
Note that the acceleration $\ddot{x}(t)$ is {\it always} positive (i.e., it is in the $+x$ direction), and since the force is constant, it reaches a maximum when the total mass of the system is at a minimum, that is at $t=t_s$ ($M(t_s)=0$).

\subsection{Initial conditions}
\label{initcond}

\noindent In order to solve the equation of motion, initial conditions are required and, as envisaged by McDonald,\cite{mcdonald} they can be deduced from the conservation law for the total momentum $p_{\rm tot}$, which we write as $p_{\rm tot} (t \leq 0) = p_{\rm tot} (t>0)$. For $t \leq 0$ the momentum is obviously zero, while for later times it results from four different contributions, namely:
\begin{itemize}
\item momentum of the tank car: \\ $p_{\rm car} = m \, \dot{x}(t)$;
\item momentum of the water moving along with the car: $p_{w_{\rm car}} = M(t) \, \dot{x}(t)$; 
\item momentum of the water acting on the car (see (\ref{pw})): $p_{w_{\rm ins}} = - \dot{M}(t) \, D$;
\item momentum of the water exiting from the car; in the time interval $\Delta t$ it is given by: \\ 
$p_{w_{\rm out}} = \Delta M(t) \cdot \dot{x}(t)$.
\end{itemize}
By introducing the mass rate (which is negative) in the last contribution, $\Delta M(t) \cdot \dot{x}(t) = - \dot{M}(t) \, \Delta t \cdot \dot{x}(t)$, the conservation law for the momentum becomes:\footnote{Strictly speaking, Eq.$\,$(\ref{momentum}) holds only for a finite time interval $\Delta t$, while the general conservation law, valid at any time $t$, should be written according to McDonald\cite{mcdonald} as:
\[
\left[m + M(t) \right] \dot{x}(t) - \dot{M}(t) \, D - \int_0^t \!\! \dot{M}(t^\prime) \, \dot{x}(t^\prime) \, \drm t^\prime = 0 \, .
\]
However, as it will be evident below, this mathematically and physically relevant point does not affect the results of the following analysis, so that we judge it is not convenient here for pre-calculus students to exhibits unnecessary mathematical complications. Anyway, for calculus-based courses for undergraduate students, such subtleties could be pointed out.}
\be
\left[m + M(t) \right] \dot{x}(t) - \dot{M}(t) \, D - \dot{M}(t) \, \Delta t \cdot \dot{x}(t) = 0
\label{momentum}
\ee
From this, for $t=0$ (and, thus, $\Delta t=0$) we deduce that the initial velocity $\dot{x}(0) = \dot{x}_0$ is given by:
\be
\dot{x}_0 = \frac{\dot{M}_0 D}{m + M_0} = - \frac{2D}{t_s} \frac{\mu}{1+\mu} \, ,
\label{v0}
\ee
where in the last equality we have conveniently introduced the physically relevant parameter (see below) $\mu = M_0/m$ denoting the ratio of the total water mass versus the tank car mass.

Interestingly enough, the tank car start to move at $x_0=0$ with a non-vanishing velocity $\dot{x}_0 \neq 0$ (for $D=0$ the car does not move due to symmetry reasons, while for $M_0=0$ no water is present in the car, and so no motion starts). Also, such initial velocity is always {\it negative}, $\dot{x}_0 < 0$, thus requiring an initial thrust on the car in the $-x$ direction, being necessarily due to the action of gravity at the initial time instant. Indeed, we can envisage the following situation. When the draining hole is set free, a water element $\Delta M$ exits from the hole downwards due to gravity; the car walls start to push  water towards the hole, the left wall pushing more water than the right one in the case in Fig.$\,$\ref{fig1}, where the hole is located in the positive $x$ direction. For the action-reaction principle, the resulting force made by water on the car walls (initial throw) then  acts in the negative $x$ direction, and thus $\dot{x}_0 < 0$. As a result, the initial throw given by gravity on the water element falling (vertically) from the hole, when this is set free, provides the (horizontal) initial throw in the $-x$ direction to the system.

\subsection{Qualitative description of the motion}

\noindent The picture thus emerged is then the following. The leaky tank car problem is just a variable mass problem ruled by a constant positive force, which is basically dictated, in the present model, by the requirement of the Bernoulli equation (that is, ultimately, by energy conservation), resulting into a positive (non constant) acceleration at any time. Momentum conservation requires, instead, that the initial velocity be negative, this paralleling the more well known example of a ball launched upwards under the action of gravity (which gives, however, a constant acceleration on Earth surface). Then, the car initially moves towards left (in Fig.$\,$\ref{fig1}), but the modulus of its velocity decreases (since $\ddot{x} > 0$), approaching zero: at this time ($t_{\rm inv}$), an inversion of motion takes place. Three possible cases may then occur, according to the actual emptying time: 1) no inversion takes place and, once the ``fuel" (the water exiting the hole) is run out, the car still moves towards left by inertia ($t_{\rm inv} > t_s$); 2) the car stops when completely empty ($t_{\rm inv} = t_s$); 3) the car undergoes an inversion of its motion and, after the emptying time is reached, it continues moving towards right ($t_{\rm inv} < t_s$). The puzzle of the inversion of the motion, already envisaged by McDonald,\cite{mcdonald} is thus easily explained physically.

Of course, by neglecting any friction, the water fallen from the car continuously moves in order to assure total momentum conservation. In case 2), it is thus clear that a part of the water fallen moves leftward, while another part moves rightward: the water fallen from the car invert itself the direction of motion before the car does. All these interesting points, now discussed only qualitatively, will be addressed appropriately below. 

\section{A more realistic physical model}

\noindent The physical input introduced above sets definitely the leaky tank car problem, also reproducing exactly the known results in the literature coming from the (mathematical) solution of the equation of motion. However, before providing such solution and discussing its physical features, it is here probably more useful -- again, in order to exploit all the physical potentialities of the problem -- to inspect a bit more the model adopted until now, which is just what introduced by McDonald.\cite{mcdonald} In such a model, as recalled above, we have assumed that water flows homogeneously inside the car from the center of mass $G$ of water to the hole, and this led to the equation of motion (\ref{eom}) with the initial condition (\ref{v0}). 

\begin{figure}[t]
\includegraphics[width=7cm]{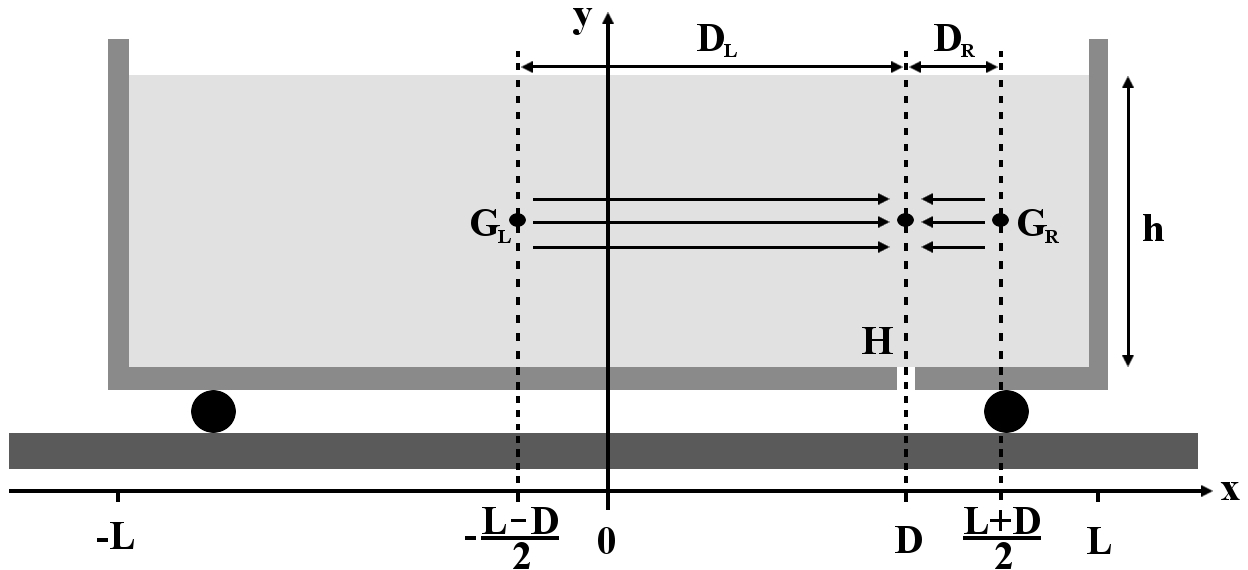}
\caption{A leaky tank car (filled with liquid at height $h$) of length $2L$ with a draining small hole $H$ at position $D$ off center: water flows from its ``left" center of mass $G_L$ to the hole and from its ``right" center of mass $G_R$ to the hole (see text).}
\label{fig2}
\end{figure}

More realistically, such overall flux in the $+x$ direction is due to water flowing rightward from the left of the hole as well as to water flowing leftward from the right of the hole. With reference to Fig.$\,$\ref{fig2}, we can then generalize the previous physical model to a more precise one by differentiating the centers of mass $G_L,G_R$ of the water at the left/right of the hole, located at distances $D_L=(L+D)/2$ and $D_R=(L-D)/2$, respectively. Of course, water flowing from $G_L,G_R$ to the hole (whose volumes will be denoted with $V_L, V_R$, respectively) is not the same since, from simple geometry, we have:
\be
V_L = \frac{D_L}{L} \, V \, ,\qquad \qquad V_R = \frac{D_R}{L} \,  V \, .
\label{volumes}
\ee
Similarly (with reference to the total mass $M=M(t)$ in the tank car), we have:
\be
M_L = \frac{D_L}{L} \, M \, , \qquad \qquad M_R = \frac{D_R}{L} \, M \, ,
\ee
respectively. This means that the momentum (\ref{pw}) of the water acting on the car and entering into Eq.$\,$(\ref{momentum}) is now given by:
\be
- \, p_{w_{\rm ins}} = \Delta M_L \cdot \frac{D_L}{\Delta t_L} - \Delta M_R \cdot \frac{D_R}{\Delta t_R} \, .
\label{pwlr}
\ee
In the McDonald model considered above, the time intervals $\Delta t_L,\Delta t_R$ for the water to cover the distances $D_L, D_R$ from $G_L,G_R$ to the hole were implicitly assumed to be equal,\footnote{Indeed, for $\Delta t_L = \Delta t_R = \Delta t$: \[ \frac{\Delta M_L}{\Delta t} \cdot D_L - \frac{\Delta M_R}{\Delta t} \cdot D_R = \frac{\dot{M}}{L} \left( D_L^2 - D_R^2 \right) = \dot{M} D \, . \]} this implying evidently that the (horizontal) velocities of the water flowing leftward and rightward were {\it different} (since $D_L \neq D_R$). However, such assumption is not very well justified, since the vertical velocities of the water falling leftward and rightward in the hole are expected to be equal, depending only on the height $h$ of the water in the car, according to the assumption A2. We can estimate such velocities in the following manner.

The quantity of water flowing in the hole rightward (see Fig.$\,$\ref{fig2}) is expected to be greater than the corresponding one flowing leftward; in order to make quantitative such statement, we can imagine that the hole is subdivided into two parts (left and right) of surface areas  $S_{hL}$ and $S_{hR}$, respectively ($S_{hL}+S_{hR}=S_{h}$). By denoting with $S_{cL}$, $S_{cR}$ the horizontal surface areas of the tank car on the left, right of the hole ($S_{cL}+S_{cR}=S_{c}$), we can reasonably assume that the quantity of water flowing in the hole from the left and from the right is proportional to such areas, that is:
\be
\frac{S_{cL}}{S_{hL}} = \frac{S_{cR}}{S_{hR}} \, .
\label{sratio}
\ee
Since the vertical velocities of water exiting the drain leftward and rightward are given, from Eq.$\,$(\ref{ce}), by 
\be
v_L \, = \, - \frac{S_{cL}}{S_{hL}} \, \dot{h} \, , \qquad \qquad 
v_R \, = \, - \frac{S_{cR}}{S_{hR}} \, \dot{h} \, ,
\ee
respectively, from Eq.$\,$(\ref{sratio}) it is then evident that $v_L = v_R$. As said above, since these vertical velocities are equal, we can expect that the horizontal velocities of the water flowing leftward and rightward are equal as well (since, in a sense, the heights $h_L = h_R = h$), and given the fact that the distances covered $D_L$ and $D_R$ are different, necessarily we have that $\Delta t_L \neq \Delta t_R$:
\be
\frac{D_L}{\Delta t_L} = \frac{D_R}{\Delta t_R} \qquad \Rightarrow \qquad \Delta t_L = \frac{D_L}{D_R} \, \Delta t_R \, .
\label{dtlr}
\ee
Moreover, we can also calculate such different time intervals from the conservation law for the volumetric flow rate ($\Delta V$ is the volume of water flowing in the time $\Delta t$, and similarly for $L,R$),
\be
\frac{\Delta V}{\Delta t} = \frac{\Delta V_L}{\Delta t_L} \, + \, \frac{\Delta V_R}{\Delta t_R} \, .
\label{cf}
\ee
From Eqs.$\,$(\ref{dtlr}), (\ref{cf}) and (\ref{volumes}), it follows immediately that
\be
\Delta t_L = \frac{2 D_L}{L} \, \Delta t \, , \qquad \qquad \Delta t_R = \frac{2 D_R}{L} \, \Delta t \, ,
\ee
or, even more interestingly,
\be
\frac{D_L}{\Delta t_L} = \frac{D_R}{\Delta t_R} = \frac{L/2}{\Delta t} \, .
\label{vlvr}
\ee
That is: the horizontal velocities with which the water flows leftward and rightward are both equal to an ``average" velocity with which water covers globally a distance $L/2$ in the time $\Delta t$.

The most interesting -- and unexpected -- consequence of such result is that what deduced above within the McDonald model still holds, provided that a providential replacement be implemented. Indeed, the horizontal force in Eq.$\,$(\ref{nl}) is now given more realistically by $F = F_L - F_R$, where $F_L, F_R$ is the horizontal force acting on the left, right wall in the system tank car + water due to the water flowing leftward, rightward from $G_L, G_R$ to the draining hole in the $\pm x$ direction. As above, $F = F_L - F_R$ is constant, but the same applies to the components $F_L, F_R$ since, by placing the draining hole on the far right (left) of the car, we will have $F=F_L$ ($F=F_R$), from which we easily deduce that $F_L$ and $F_R$ are constant as well. The momentum in Eq.$\,$(\ref{pw}) is now replaced by 
\be
p_{w_{\rm ins}^L} = - \dot{M} \, \frac{D_L}{2} \, , \qquad \qquad p_{w_{\rm ins}^R} = - \dot{M} \, \frac{D_R}{2} \, ,
\label{pwlr2}
\ee
while the impulse-momentum theorem applied separately to $F_L$ and $F_R$ gives:
\be
F_L = \frac{M_0 (L+D)}{2 t_s^2} \, , \qquad \qquad F_R = \frac{M_0 (L-D)}{2 t_s^2} \, .
\label{flfr}
\ee
As expected, $F_L$ is greater than $F_R$ and, more intriguingly:
\be
F = F_L - F_R = - \, \frac{\dot{M}_0 D}{2 t_s} = \frac{M_0 D}{t_s^2} = \frac{\ddot{M} D}{2} \, ,
\label{forcelr}
\ee
which result is just half of that in Eq.$\,$(\ref{force}). This means that the equation of motion in (\ref{eom}) is now replaced by:
\be
\left[ m + M(t) \right] \ddot{x}(t) = \frac{M_0 D}{t_s^2} \, ,
\label{eomlr}
\ee
while the conservation law for the momentum (\ref{momentum}) writes:
\be
\left[m + M(t) \right] \dot{x}(t) - \, \frac{\dot{M}(t) \, D}{2} \, - \dot{M}(t) \, \Delta t \cdot \dot{x}(t) = 0 \, ,
\label{momentumlr}
\ee
from which the initial condition for the velocity is deduced:
\be
\dot{x}_0 = \frac{\dot{M}_0 D}{2(m + M_0)} = - \frac{D}{t_s} \frac{\mu}{1+\mu} \, .
\label{v0lr}
\ee

\section{Motion of the leaky tank car}

\noindent The motion of the system considered is obviously given by the solution of Eq.$\,$(\ref{eomlr}), with $M(t)$ given by (\ref{mt}), for a tank car starting its motion at $x_0=0$ with initial velocity given by (\ref{v0lr}). Differently from what done until now, however, this requires necessarily the use of calculus, that is how to solve a differential equation with non-constant coefficients. For undergraduate students, this is not a serious problem, but for pre-calculus ones the following expressions should be provided them without too much pain.

\begin{figure}[t]
\bt{c}
\includegraphics[width=8.5cm]{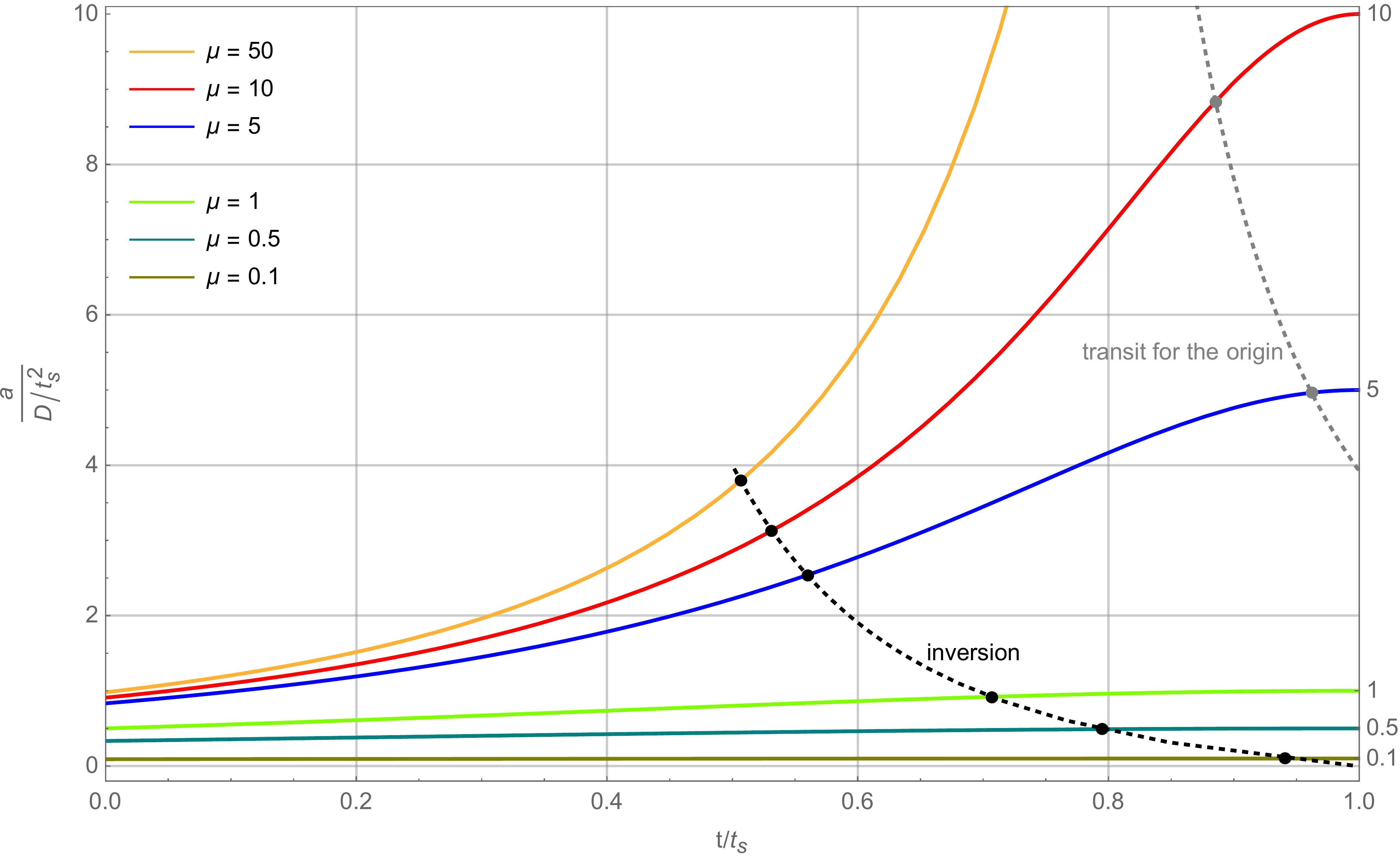} \\
\includegraphics[width=8.5cm]{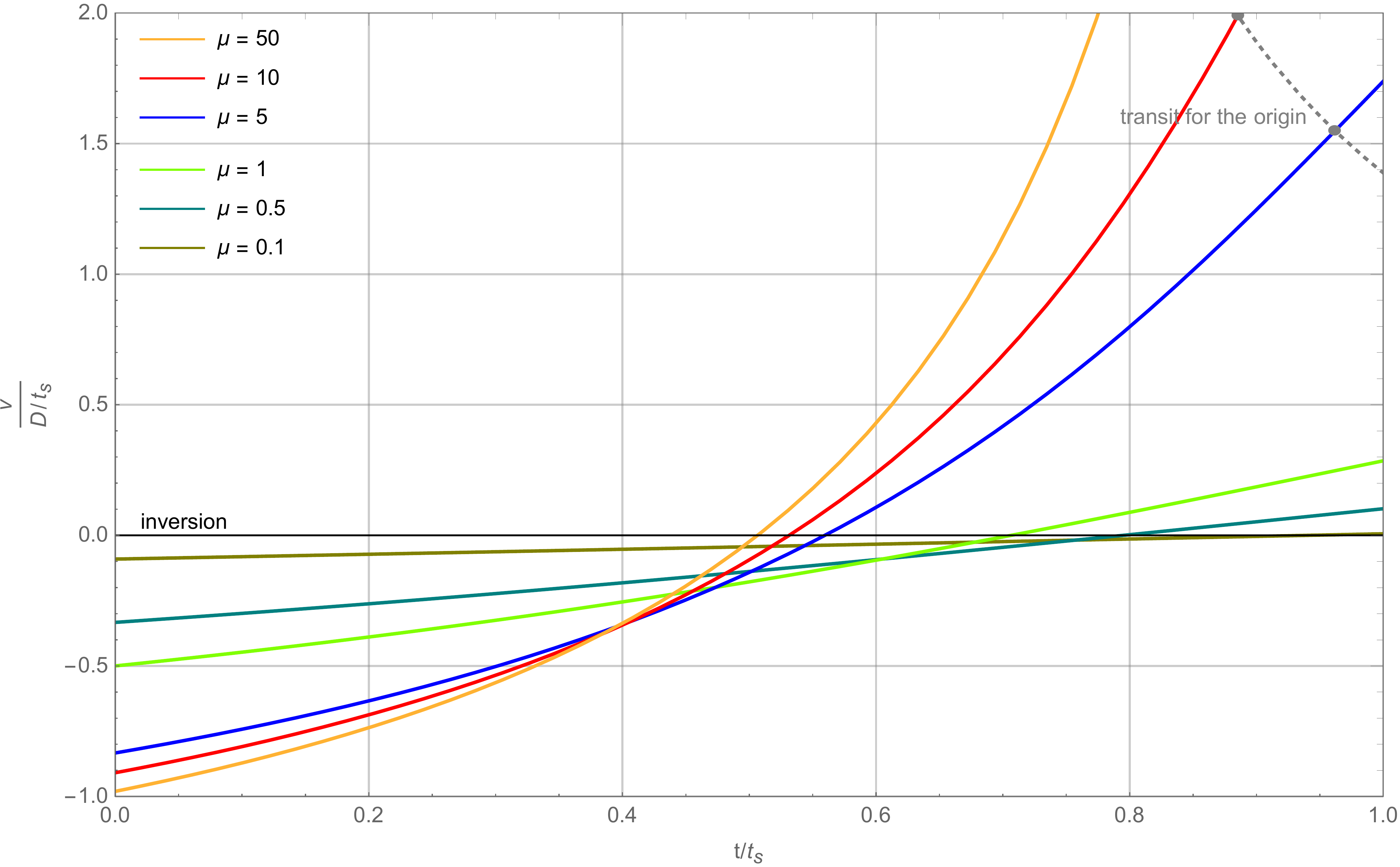} \\
\includegraphics[width=8.5cm]{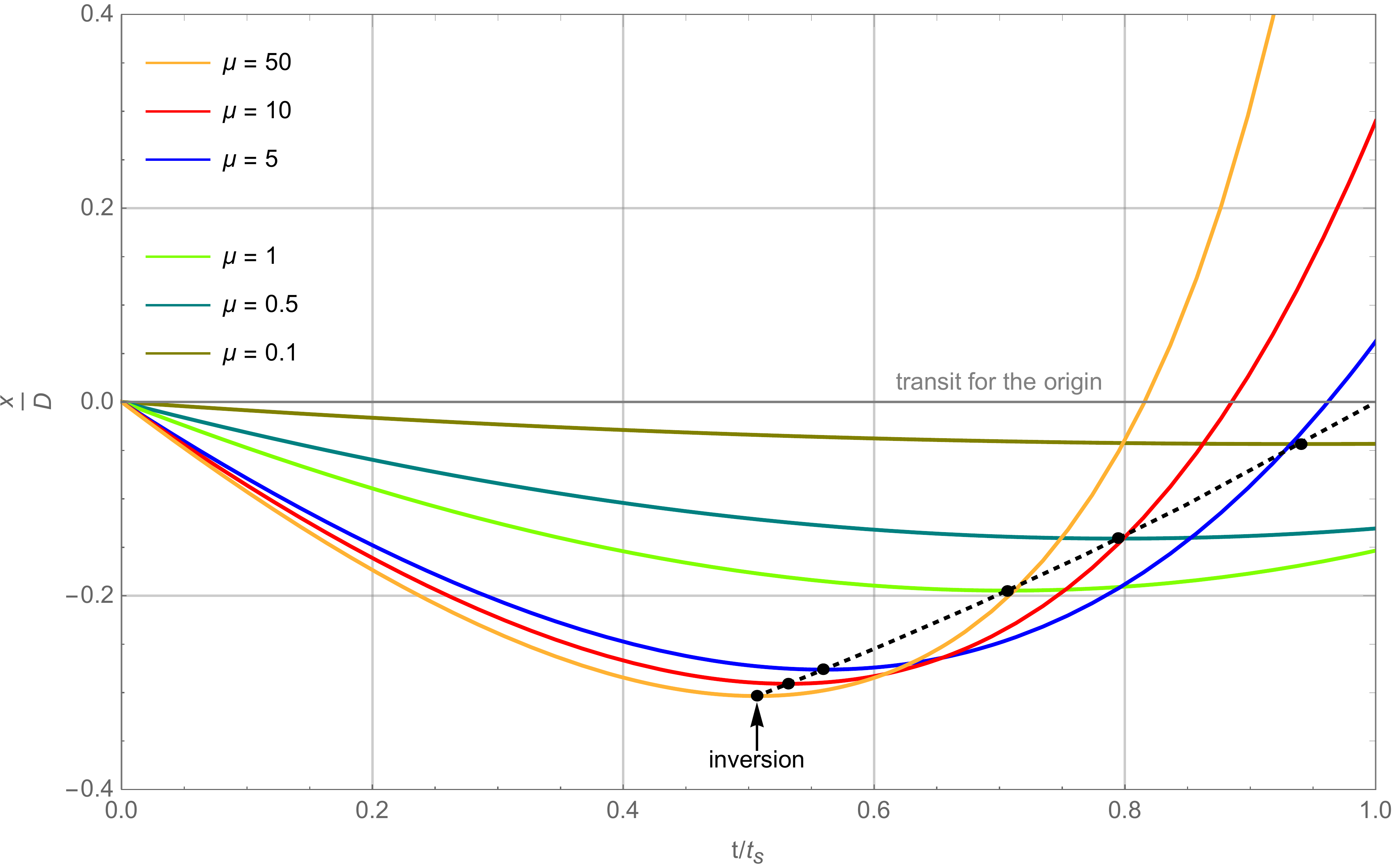} 
\et
\caption{Acceleration $a$, velocity $v$ and position $x$ of a leaky tank car as function of time, for different values of the mass ratio parameter $\mu$. The inversion of the motion and the transit of the car for the origin (when realized) are pointed out.}
\label{fig3}
\end{figure}

\subsection{Solution of the equation of motion}

\noindent Such solution\cite{mcdonald} for acceleration $a(t)=\ddot{x}(t)$, velocity $v(t)= \dot{x}(t)$ and position $x(t)$ can be cast conveniently in a physically meaningful (though quite involved) form as follows:
\begin{align}
a(t) &= \frac{D}{t_s^2} \, \frac{\mu}{\dis 1 + \mu \left( 1 - \frac{t}{t_s} \right)^2} \, ,\label{aeq}
\\
v(t) &= - \, \frac{D}{t_s} \, \frac{\mu}{1 + \mu} \left\{ 1 - \frac{1 + \mu}{\sqrt{\mu}} \left[ \arctan \sqrt{\mu} \vphantom{\left( 1 - \frac{t}{t_s} \right)} \right. \right. \nonumber \\ &  \quad \left. \left. - \arctan \sqrt{\mu} \left( 1 - \frac{t}{t_s} \right) \right] \right\} , \label{veq} 
\\
x(t) &= - \, D \left\{ \frac{\mu}{1 + \mu} \, \frac{t}{t_s} - \frac{1}{2} \left[ \log \frac{1 + \mu}{\dis 1 + \mu \left( 1 - \frac{t}{t_s} \right)^2} \right. \right. \nonumber \\ &  \quad - 2 \sqrt{\mu} \left( 1 - \frac{t}{t_s} \right) \left( \arctan \sqrt{\mu} \vphantom{\left( 1 - \frac{t}{t_s} \right)} \right. \nonumber \\ &  \qquad \left.  \left. \left. - \arctan \sqrt{\mu} \left( 1 - \frac{t}{t_s} \right) \right) \right] \right\} . \label{xeq} 
\end{align}
Note that, of course, all such quantities are proportional to $D$, so that they vanish for a central drain hole at $D=0$. Also, the scales for position, velocity and acceleration are set by $D$, $D/t_s$ and $D/t_s^2$, respectively (while $t_s$ is the appropriate scale for time). Indeed, distance $D$ of the hole off the center and emptying time $t_s$ being the only physical kinematic quantities describing the problem. Instead, the only dynamic quantity relevant for the system is, as already pointed out, the mass ratio $\mu = M_0/m$, and in Fig.$\,$\ref{fig3} we plot the above solutions for different values of such parameter from the starting at $t=0$ to the emptying of the tank car at $t=t_s$. The initial and final values of $x,v,a$ are given by:
\bea
& \!\! \!\! \!\!  \!\! & \left\{ \ba{l}
\dis x(t=0) = 0 , \\ 
\dis v(t=0) = - \, \frac{D}{t_s} \,  \frac{\mu}{1 + \mu} \, ,   \\
\dis a(t=0) = \frac{D}{t_s^2} \,  \frac{\mu}{1 + \mu} \, ,  \\
\ea \right. \label{inavx} \\
&  \!\! \!\! \!\!  \!\! & \left\{ \ba{l}
\dis x(t=t_s) = D \left[ \frac{1}{2} \log \left( 1 + \mu \right) - \frac{\mu}{1 + \mu} \right] ,  \\
\dis v(t=t_s) = - \, \frac{D}{t_s} \left(  \frac{\mu}{1 + \mu} - \sqrt{\mu} \, \arctan \sqrt{\mu} \right) ,  \\
\dis a(t=t_s) = \frac{D}{t_s^2} \,  \mu \, , 
\ea \right. \label{finavx}
\eea
respectively. 

In these figures, we have highlighted the physically interesting points where the car undergoes an inversion of its motion (for different values of $\mu$), which is evidently defined by the equation $v(t_{\rm inv}) =0$, that is (for finite values of $M_0, m$):
\be
t_{\rm inv} = t_s \, \frac{\dis \frac{1 + \mu}{\sqrt{\mu}} \, \tan  \frac{\sqrt{\mu}}{1 + \mu}}{\dis
1 + \sqrt{\mu} \, \tan  \frac{\sqrt{\mu}}{1 + \mu}} \, .
\label{tinv}
\ee
The realization of such inversion obviously depends on the final values in (\ref{finavx}), that is on the emptying time and the mass ratio parameter $\mu$, and {from Fig.$\,$\ref{fig3} we see that decreasing $\mu$ leads to a later inversion.}

Once inversion of motion is realized, one can also ask if the car is able to transit again for the origin\footnote{This point was raised by McDonald,\cite{mcdonald} but we find it not particularly intriguing.} {\it before} it empties (surely, after inversion, the car will transit for the origin by inertia, if we assume no friction); of course, as arguable from Fig.$\,$\ref{fig3}, this occurs only for sufficiently large values of $\mu$.

\subsection{Limiting cases}
\label{limcas}

\noindent One may wonder if the involved mathematical expressions in Eqs.$\,$(\ref{aeq}-\ref{xeq}) may assume more physically appealing forms in some limiting cases, namely for very large and small $\mu$ values, which for undergraduate student employing calculus may result into useful exercises about Taylor expansions (for $\mu \rightarrow \infty$, a novel parameter $\nu = 1/\mu$ tending to zero has to be introduced). The following results may then be appreciated only in a calculus-based presentation, although some physical cases could be discussed as well within a pre-calculus presentation. For technical details, see Appendix \ref{app1}.

Interesting enough, and apparently unexpected, such limiting cases are not ruled {\it solely} by the mass ratio parameter $\mu$, but rather separately by the $M_0$ and $m$ mass parameters. The reason is that a hidden dependence on $M_0$ is present in the emptying time $t_s$ that, from Eq.$\,$(\ref{ts}), we can now write as:
\be
t_s \, = \, \sqrt{\frac{2 S_c M_0}{\rho S_h^2 g}} \, .
\label{ts2}
\ee
For an appreciably non-vanishing water mass ($M_0 \neq 0$), the emptying time $t_s$ is finite, and standard Taylor expansion methods may be applied to Eqs.$\,$(\ref{aeq}-\ref{xeq}), although with some caveats (see Appendix \ref{app1}). 

For large car masses ($m \rightarrow \infty$; $\mu \rightarrow 0$), the leaky tank car just undergoes an approximately uniformly accelerated motion:
\be
\ba{c}
\dis a \simeq \frac{D}{t_s^2} \, \mu \, , \quad
v \simeq - \, \frac{D}{t_s} \left( 1 - \frac{t}{t_s} \right) \mu \, , \\ \\
\dis x \simeq - D  \left( 1 - \frac{t}{2t_s} \right) \frac{t}{t_s} \, \mu \, ,
\ea
\ee
ending at $t=t_s$ {with vanishing velocity.} Note, however, that in such case, the small quantity of water in the car, when compared to the mass of the car, does not allow an effectively appreciable motion: the (approximately constant) acceleration is very small as well as the initial velocity $v(t\!=\!0) \!=\! - (D/t_s) \mu$, so that the car moves effectively just for a negligible distance, from $x(t\!=\!0)=0$ to $x(t\!=\!t_s) \!=\! - D \mu /2 \rightarrow 0$.

For small car masses ($m \rightarrow 0$; $\mu \rightarrow \infty$), instead, no appreciable inertia is opposed by the car to its motion, which is then completely dominated by the action of the moving water. The car effectively covers a large distance (tending to infinity for $\mu \rightarrow \infty$) before emptying and, since $t_s$ is finite, acceleration as well as velocity increase considerably when the emptying time is approaching. Then, {the inversion of motion occurs at a time approximately equal to half the emptying time:}
\be
t_{\rm inv} \simeq \left( \frac{1}{2} + \frac{1}{3 \mu} \right) t_s \,
\ee
as can be deduced by Taylor expanding Eq.$\,$(\ref{tinv}) around $\nu = 1/\mu = 0$. Just until this time (or, more accurately, just at the beginning, for $t \ll t_s$), the acceleration can be approximated to a constant equal to $D/t_s^2$ (see Appendix \ref{app1}).

In the limit of vanishing or infinite water mass, the approximate analysis becomes a bit more tricky (and we refer the reader to Appendix \ref{app1}), since the emptying time $t_s$ is no more finite, and a different time scale rules the motion for a finite car mass ($m \neq 0$).

For vanishing water masses ($M_0 \rightarrow 0$; $\mu \rightarrow 0$), the emptying process rapidly reaches an end, and during such negligibly small time the acceleration may be considered approximately constant. {However, the velocity is as well exceedingly small, so that the car is only able to cover a very small distance during its motion.}

{Just the opposite happens for very large water masses ($M_0 \rightarrow \infty$; $\mu \rightarrow \infty$), for which the emptying process lasts for an exceedingly large time ($t_s \rightarrow \infty$).} Even in this case we have approximately a uniformly accelerated motion, but now both acceleration and velocity practically vanish due to the extremely large inertia suffered by the system, the car again covering only a negligibly small distance.

\section{Motion of the water flowing from the car}

\noindent The motion of the tank car when water drains through the hole, as described above, is not the only interesting problem to study for such a physical system. Indeed, it is as well quite intriguing to learn also about the (horizontal) motion of the water exiting the car, and this for several reasons, primarily originated by asking what happens to water flowing from the car when inversion occurs. 

In the following, we will consider the problem in some detail but, without embarking in complicate hydrodynamics descriptions (which are certainly untimely in the present framework), once instead completely known the motion of the tank car, we can fruitfully make recourse to the momentum conservation law in order to obtain some insight into the motion of the water exiting the car.

Here, the problem is to estimate the four different momentum contributions discussed in Sect. \ref{initcond}, while determinate solutions (\ref{aeq}-\ref{xeq}) (with Eqs.$\,$(\ref{mt}),(\ref{mpt})) apply for different values of the mass ratio parameter $\mu$ (including, in case, also the limiting cases already considered above). This is apparently just a useful mathematical exercise (for students trained in calculus), which we  indeed address in Appendix \ref{app2}, but the physical interpretation of the results obtained, as well as the conclusions drawn from it, are surprisingly interesting. We then dwell here upon the physical aspects of this problem, referring the reader to Appendix \ref{app2} for technical details.

The focus is, evidently, on the contribution $p_{w_{\rm out}}$ to the total momentum of the system introduced above. Of course, for $t=0$ we have $p_{w_{\rm out}} =0$, since no momentum is carried by the water exiting the car at the beginning (for $t=0$ we still have no water flowing horizontally outside the car). Afterward, $p_{w_{\rm out}}$ will be negative: this is a direct consequence of the fact that the car start to move in the $-x$ direction (see above) -- until inversion occurs (if any) -- and that the water exiting the car follows such motion, just as a projectile released by a plane. For $t= t_{\rm inv}$, $p_{w_{\rm out}}$ takes a minimum, since from now on the car -- inverting its motion -- releases water from the hole in the $+x$ direction, thus increasing the value of $p_{w_{\rm out}}$, while still negative. This behavior is, indeed, confirmed by the calculations (and plots) reported in Appendix \ref{app2}, from which however we deduce that, for finite $m, M_0$, when the car is empty we still have $p_{w_{\rm out}} (t_s) < 0$. This is not a surprising result, though naively unexpected, and does not contradict the obvious fact that the last drops of water exiting the car do follow the motion of the car in the $+x$ direction (once inversion occurred). It simply means that the {\it overall} motion of the water exited from the car is {\it always} in the $-x$ direction, even after inversion. 

This is due to the fact that, according to Eq.$\,$(\ref{mt}), most of the water exits the car before inversion, as can be easily seen by evaluating the mass $M_\pm (t)$ of water exited from the car and flowing in the $\pm x$ direction as function of time:
\begin{align}
M_-(t) &= \left\{ \ba{ll} M_0 - M(t) , \qquad & 0 \leq t \leq t_{\rm inv} , \\ M_0 - M_{\rm inv} , \qquad & t > t_{\rm inv} , \ea \right.
\label{mtm} \\
M_+(t) &= \left\{ \ba{ll} 0 , \qquad & 0 \leq t \leq t_{\rm inv} , \\ M_{\rm inv} - M(t) , \qquad & t_{\rm inv} < t \leq t_s , \\ M_{\rm inv} , \qquad & t> t_s \ea \right.
\label{mtp}
\end{align}
($M_{\rm inv} \equiv M(t_{\rm inv})$).\footnote{It is a useful exercise for students to deduce the expressions in Eqs.$\,$(\ref{mtm}), (\ref{mtp}), also giving the following hints: 1) $M(t) + M_+(t) + M_-(t) = M_0$ for any $t$; 2) $M_+(t) + M(t) = M_{\rm inv}$ for $t_{\rm inv} \leq t \leq t_s$; 3) $M_+(t) =0$ for $t \leq t_{\rm inv}$.} The large quantity of water still flowing by inertia in the $-x$ direction after inversion, notwithstanding the additional water, with minor mass, starting to flow in the $+x$ direction, thus explains the overall negative value of $p_{w_{\rm out}}$. 

{We then remain with the following picture. At the beginning, the tank car moves in the $-x$ direction -- and so does the water flowing outside from it -- until inversion, which always takes place for any value of $\mu$: indeed, from the analysis of the water flowing outside the car, if the inversion of motion would not occur, there would be nothing going in the $+x$ direction assuring momentum conservation. After the inversion, the car moves in the $+x$ direction, the water still draining from it doing the same, while the water already exited from the car continues its flowing in the $-x$ direction. Such a picture is allowed by momentum conservation, since the momentum of the system (tank car) + (water inside moving with it) + (water drained from the car and moving outside) is balanced by the momentum of the water moving inside the car, which always flows globally from the center of mass towards the hole, that is in the $+x$ direction, without undergoing any inversion. While time goes by, however, such momentum decreases more and more (with the decreasing of the mass of water still inside the car), until vanishing when the emptying process is completed: since the water already exited from the car moves by inertia in the $-x$ direction, necessarily at a certain time the water has to start to drain in the $+x$ direction, this explaining the motion inversion even for water.}

\begin{figure}[t]
\includegraphics[width=8.5cm]{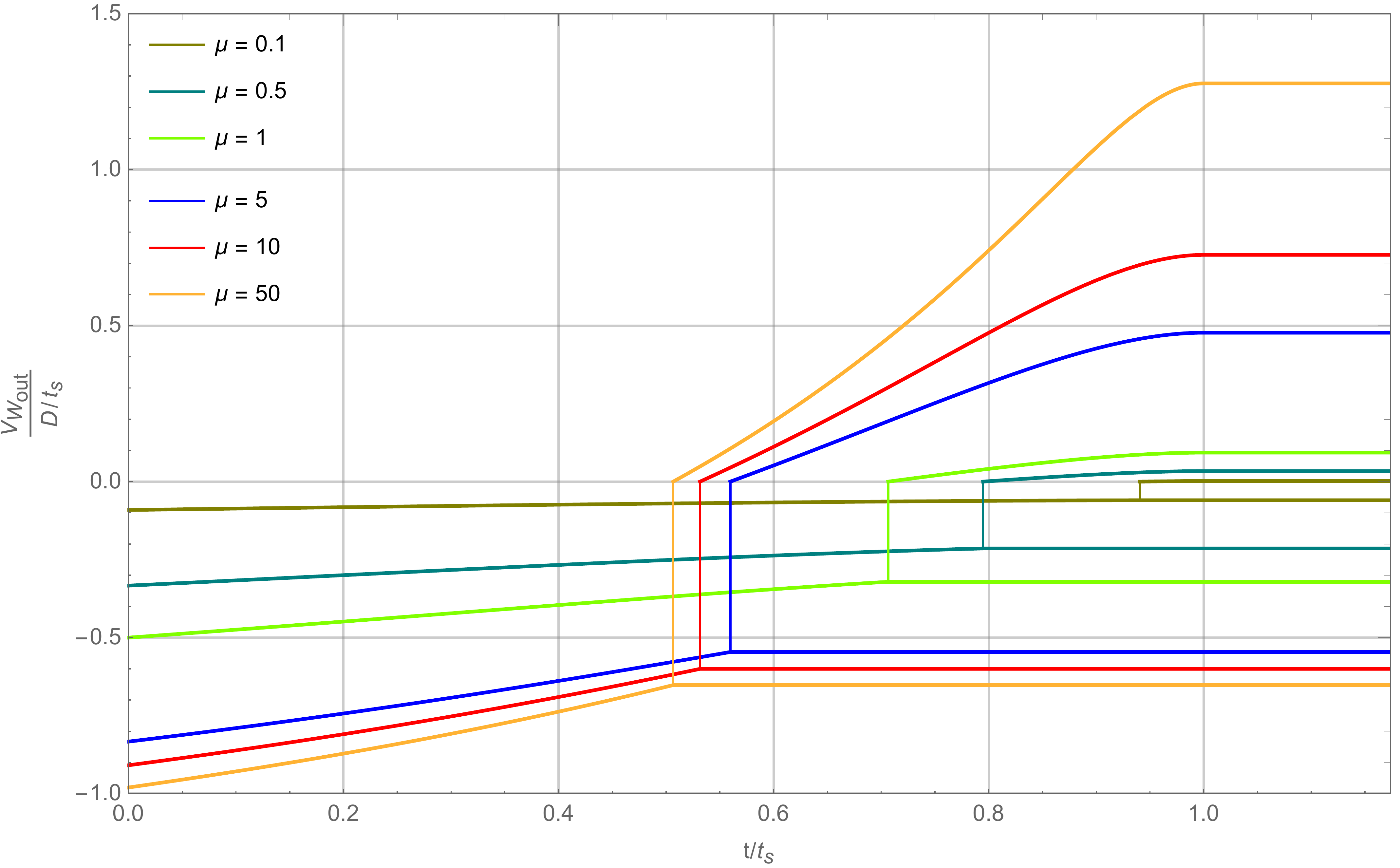}
\caption{Velocity of the water flowing outside the tank car as function of time, for different values of the mass ratio parameter. The inversion point marks the transition from a one-valued behavior (water flowing only leftward) to a two-valued one (water flowing partly leftward and rightward). Note that the final values of the rightward velocity are always greater than the corresponding values (in modulus) of the leftward velocity due to momentum conservation, since the opposite applies for the corresponding masses. By increasing the initial water mass (i.e. increasing $\mu$), final values of these velocity (in modulus), as well as the initial ones, always increase.}
\label{fig4}
\end{figure}

All this is made quantitative in Appendix \ref{app2}, and reported in the corresponding plots, where it is evident that the scale for any momentum contribution is set by the quantity $M_0 D / t_s$, as expected. Also, from those plots, it can be appreciated the dependence of $p_{w_{\rm out}}$ from the mass ratio parameter $\mu$: obviously, $p_{w_{\rm out}}$ becomes more and more negative for increasing $\mu$ (since the quantity of water mass available increases), but this is no longer true after inversion for higher values of $\mu$ (around $\mu =5$). Indeed, for higher values of the water mass, inversion takes place earlier, and more water flows outside the car in the $+x$ direction. 

Particularly interesting is the limiting case of $\mu \rightarrow \infty$ realized when the car mass is negligible compared to water mass ($m \rightarrow 0$),\footnote{For the other three limiting cases discussed above in Sect. \ref{limcas}, the water flowing outside the car just follows the fate of the car motion, presenting no particular interest.} since this is the only case when $p_{w_{\rm out}}$ again vanishes (after the starting) at the emptying time. Here $p_{w_{\rm out}}$ is totally symmetric around the inversion point that now occurs at $t_{\rm inv} = t_s/2$, being indeed described by a parabola intersecting the time axis at $t=0$ and $t=t_s$ and with its vertex at $t=t_s/2$. The tank car stops its motion when empty, and $p_{w_{\rm out}}$ vanishes at this point, but the water already exited obviously doesn't stop moving, a given (greater) quantity of water flowing (slower) in the $-x$ direction, and another (smaller) quantity flowing (faster) instead in the $+x$ direction, perfectly balancing their momenta: more precisely, $3/4 \, M_0$ water moves in the $-x$ direction with velocity $- 2/3 \, D/t_s$, while the remaining $1/4 \, M_0$ water moves in the $+x$ direction with velocity $2 \, D/t_s$, each component carrying exactly half of the momentum $p_{w_{\rm out}}$.

From the mathematical results of Appendix \ref{app2} concerning $p_{w_{\rm out}}$, it can be also deduced the behavior of the velocity $v_\pm (t)$ of water flowing leftward/rightward outside the car as function of time,\footnote{For definiteness: \[ v_\pm (t) = \frac{p^\pm_{w_{\rm out}}(t)}{M_\pm(t)} \, . \]} before and after inversion. For reference, the result is reported, for different values of the mass ratio parameter $\mu$, in Fig.$\,$\ref{fig4}, which well summarizes what described above.

\section{Generalizations}

\noindent The problem studied above is worth of intriguing generalizations, whose didactic and physical interest rises little doubt. First of all, note that we can refer to it properly as a one-dimensional (1D) problem, since the car is able to move only in the $x$ direction and, as evident from Figs.$\,$\ref{fig1} or \ref{fig2}, only any motion of water in this direction (effectively induced by motion in the vertical $z$ direction) is relevant, while what happens in the $y$ direction (not shown in those figures, indeed) doesn't matter.

\begin{figure}[t]
\bt{cc}
a1) \includegraphics[width=3.7cm]{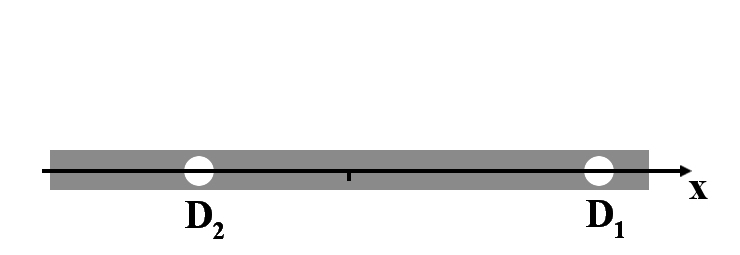} & a2) \includegraphics[width=3.7cm]{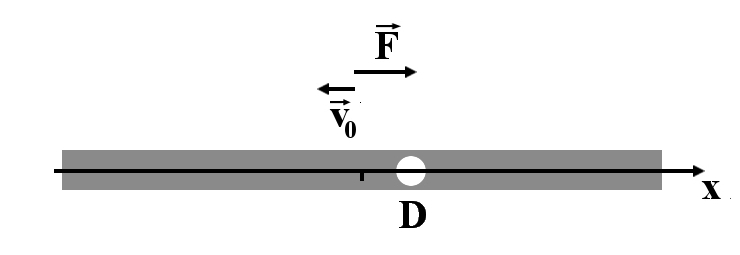} \\
b1) \includegraphics[width=3.7cm]{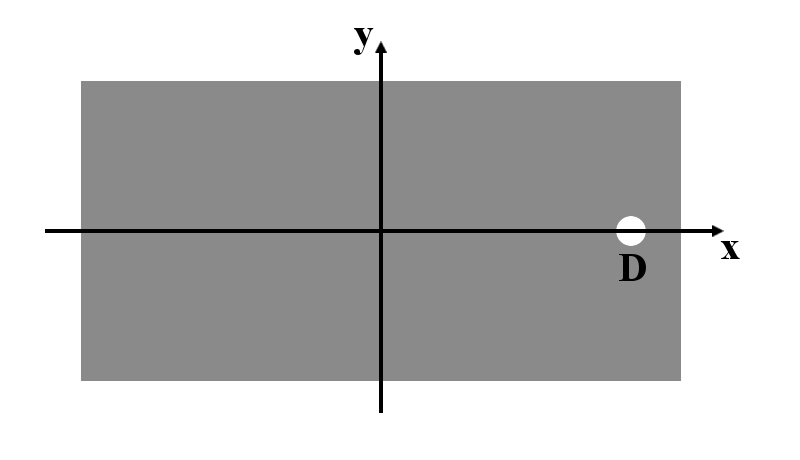} & b2) \includegraphics[width=3.7cm]{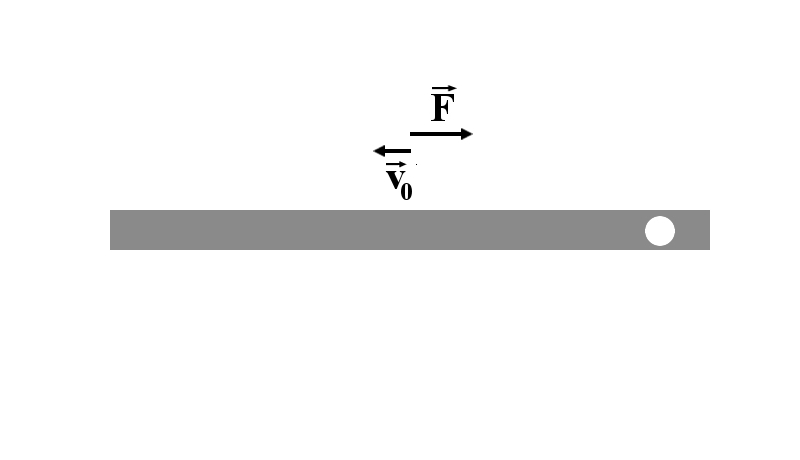} \\
c1) \includegraphics[width=3.7cm]{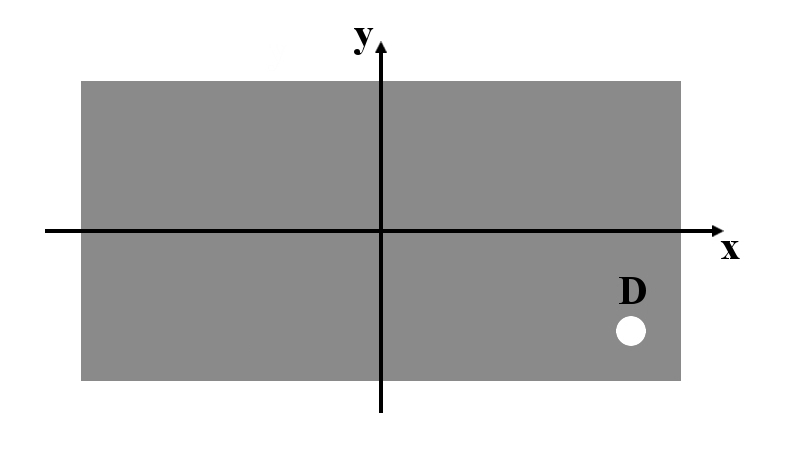} & c2) \includegraphics[width=3.7cm]{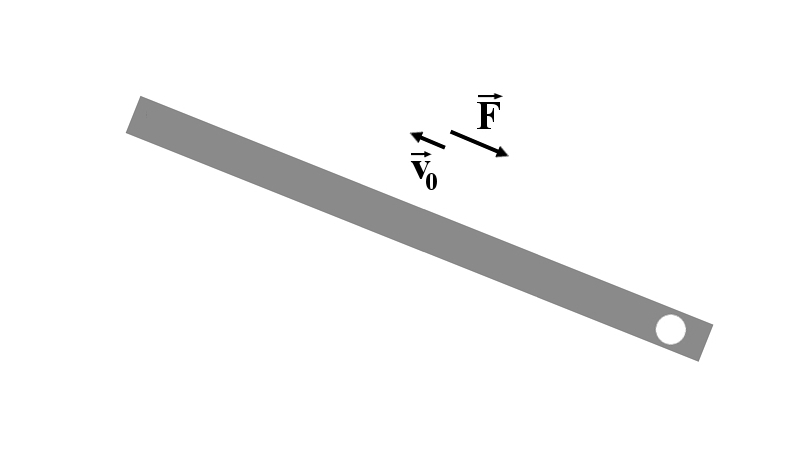} \\
d1) \includegraphics[width=3.7cm]{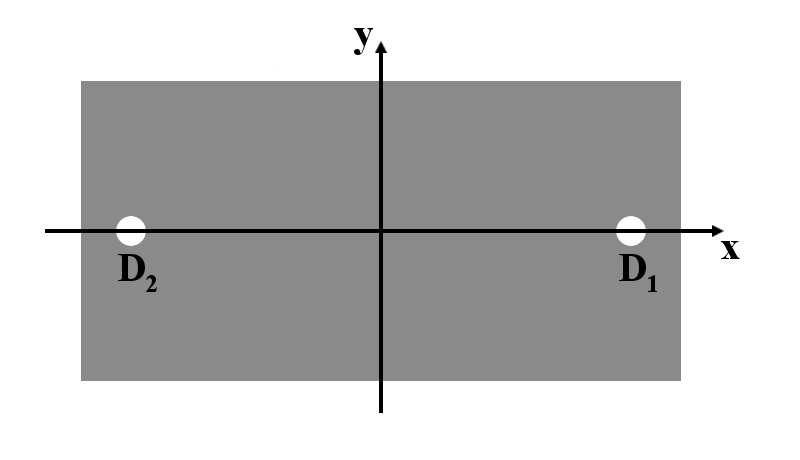} & d2) \includegraphics[width=3.7cm]{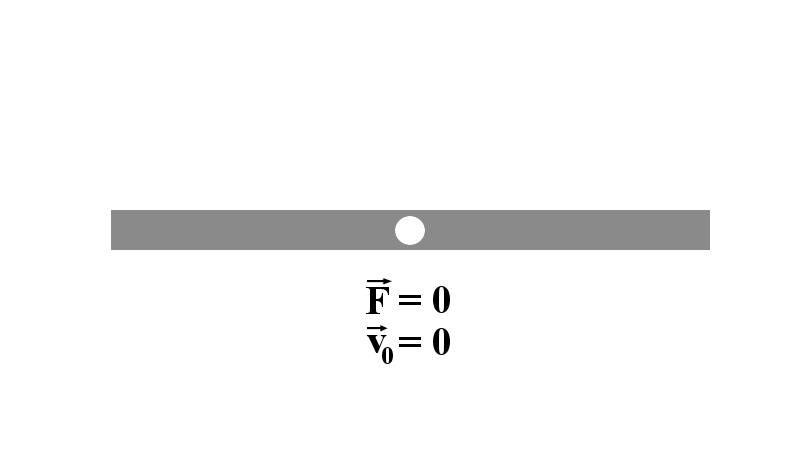} \\
e1) \includegraphics[width=3.7cm]{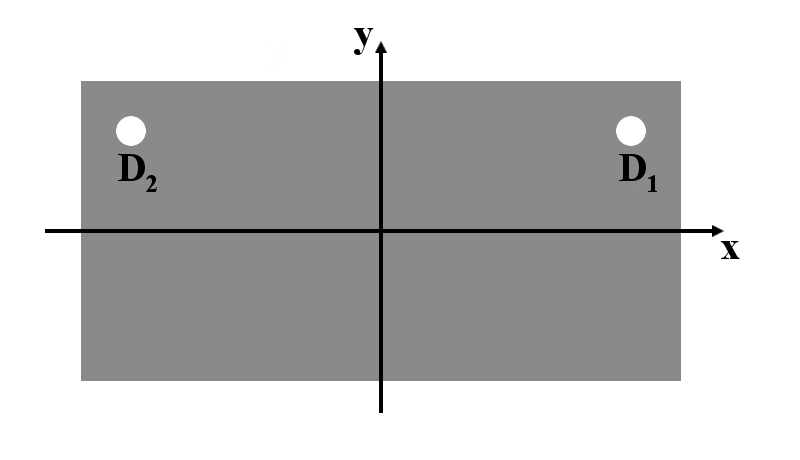} & e2) \includegraphics[width=3.7cm]{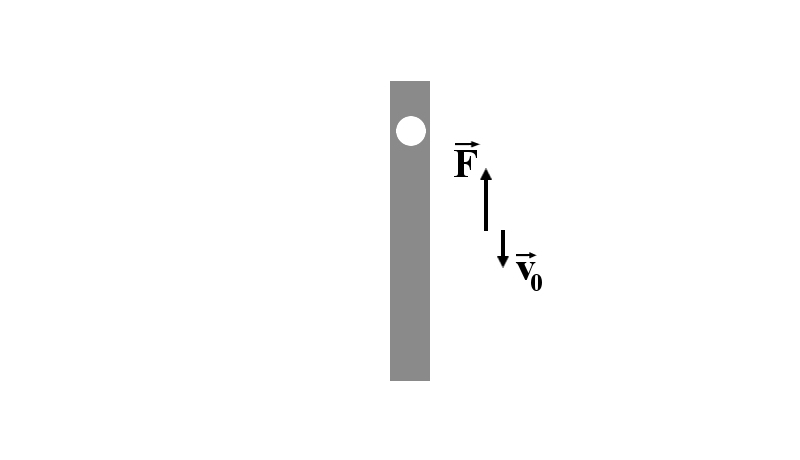} \\
f1) \includegraphics[width=3.7cm]{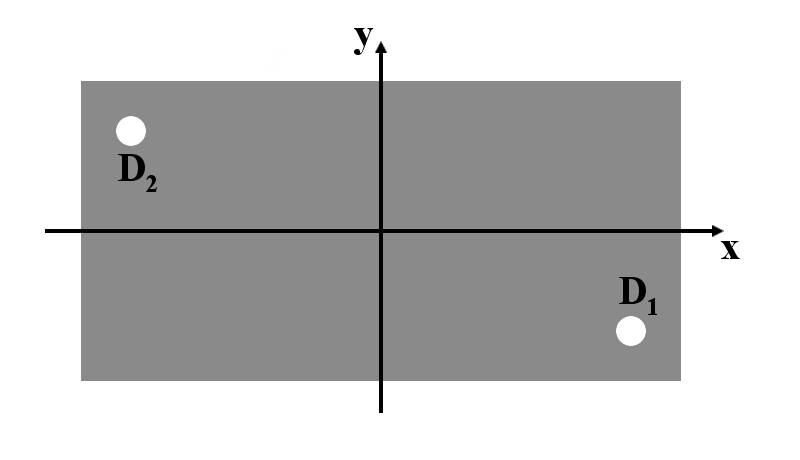} & f2) \includegraphics[width=3.7cm]{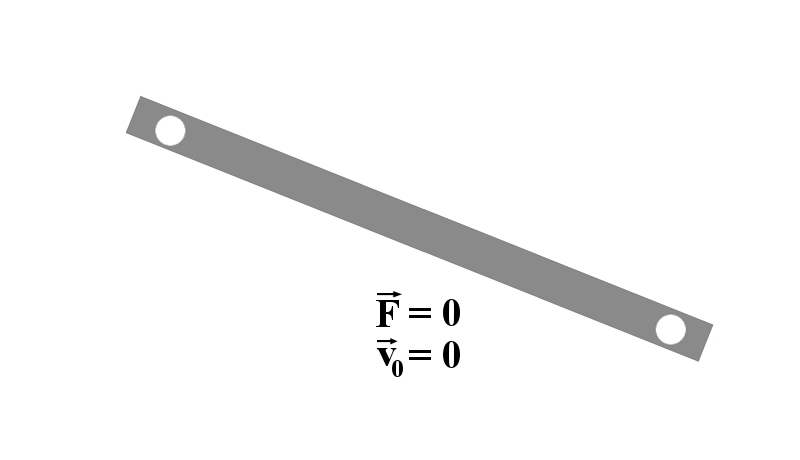} 
\et
\caption{Generalizations of the leaky tank car problem (top view). a) 1D car with two draining holes. b), c) 2D car with one hole located in different position. d), e), f)  2D car with two holes located in different positions. Left: original system. Right: equivalent 1D system.}
\label{fig5}
\end{figure}

\subsection{Two or more holes}

\noindent An immediate generalization is to consider a tank car with two (or more) draining holes, obviously located at different distances $D_1$ and $D_2$ from the center, as for example in Fig.$\,$\ref{fig5}a1 (differently from Figs.$\,$\ref{fig1},\ref{fig2}, in these figures we show a top view of the tank car, in the $xy$ plane). Of course, we now have two different forces $F_1$, $F_2 $ acting on the car, whose magnitudes -- by assuming that each hole drains just half of the water,\footnote{Here and in the following, unless otherwise specified, we consider holes with equal area, while their form is irrelevant.} of mass $M_0/2$ -- are given by:
\be
F_1 = \frac{M_0 D_1}{2 \, t_s^2} \, , \qquad F_2 = \frac{M_0 D_2}{2 \, t_s^2} \, ,
\ee
respectively. With reference to the case shown in Fig.$\,$\ref{fig5}a1, such forces act in opposite directions, so that the total force on the car, replacing that in Eq.$\,$(\ref{forcelr}), is:
\be
F = \frac{M_0 (D_1 - D_2)}{2 \, t_s^2} \, ,
\ee
and similarly we have an initial condition on the velocity in Eq.$\,$(\ref{v0lr}) with $D$ replaced by $(D_1-D_2)/2$. Of course, a leaky tank car with two holes at equal distances from the center experiences no net force, so that it doesn't move. More in general, note that 
\be
\frac{D_1 - D_2}{2} = \frac{D_1 + (- D_2)}{2} = \frac{x_{D_1} + x_{D_2}}{2} \equiv D 
\ee
(with $x_{D_1}$, $x_{D_2}$ the $x$ coordinates of the two holes) or: when assuming equal surface areas of the holes, the two holes problem is equivalent to that with just one draining hole located in the center of mass of the two holes (see Fig.$\,$\ref{fig5}a2). 

The same result applies with more than two holes. 

\subsection{Two-dimensional problems}

\noindent Case b1) of Fig.$\,$\ref{fig5}, where the (single) hole is located on the midline of the car base, shows an apparently two-dimensional (2D) problem but, due to the symmetric action of the water along the $y$ axis (resulting into a vanishing $y$-component net force), it is effectively again reducible to a 1D problem (see Fig.$\,$\ref{fig5}b2). 

The situation is instead different for the case c1), where the hole is distant $D_x$,$D_y$ from the $x$,$y$ axes. Here, both the components of the total force acting on the car are non-zero ($F_x$ pointing in the $+x$ directions, while $F_y$ in the $-y$ one), being given by:
\be
F_x = \frac{M_0 \, D_x}{t_s^2} \, , \qquad F_y = \frac{M_0 \, D_y}{t_s^2} \, .
\ee
Note that $F_x/D_x = F_y/D_y$, so that the net force $F$ always acts along the line joining the center of the car with the hole. Since a similar conclusion holds as well for the initial velocity, again the problem is equivalent to a 1D problem (\ref{forcelr}), (\ref{v0lr}) with $D=\sqrt{D_x^2 + D_y^2}$, the car moving in the center-hole direction, as in Fig.$\,$\ref{fig5}c2.

Interesting enough, analogous results apply for 2D problems with two or more holes, few typical examples being reported in Fig.$\,$\ref{fig5}d-f. Indeed, when $D_{2x} = -D_{1x}$ and $D_{2y}=D_{1y}=0$ (case d) or $D_{2y}=-D_{1y}=0$ (case f), we have $F_{2x} = - F_{1x}$ and $F_{2y}=F_{1y}=0$ (case d) or $F_{2y}=-F_{1y}=0$ (case f), and hence no net force in both cases, so the car doesn't move at all.  Instead, for $D_{2x} = -D_{1x}$ and $D_{2y}=D_{1y} \neq 0$ (case e), the problem can be reduced to that of a leaky tank car moving in the $y$ direction, as depicted in Fig.$\,$\ref{fig5}e2, a ``center of mass theorem" for the holes again being applicable (with $D=(D_{1y}+D_{2y})/2=D_{1y}$). Similar results hold for geometries different from those considered in Fig.$\,$\ref{fig5}, as the interested student can be easily check.

Here, however, another intriguing result emerges, namely that draining water from a tank car can only produce a translational motion, while {\it no rotation} can be induced. This is a consequence of the fact that, as envisaged above, the net force $F$ acting on the car always points in the direction of the line joining the center of mass of the car with the given hole, which in turn is a direct physical consequence of the fact that water effectively moves from the center of mass to the holes. As a result, no torque acts on the car at all, and the angular momentum conservation law applies not only to the car+water system, but as well to the water and -- especially -- to the car separately. If initially the car does not spin, water drainage never induces rotation (and, anyway, its angular velocity does not change). Interestingly, the leaky tank car problem is pure translational. 

\subsection{A center of mass theorem}

\noindent Although already anticipated above, it is instructive to give an explicit statement of the ``center of mass theorem", whose formal proof is straightforward from what discussed above: 
\begin{quote}
A leaky tank car problem with $n$ holes of equal area, located at $(D_{1x}, D_{1y})$, $(D_{2x}, D_{2y})$, \dots $(D_{nx}, D_{ny})$ is equivalent to a problem with just one hole located at $(D_{x}, D_{y})$, with
\be
D_x = \sum_{i=1}^n \frac{D_{ix}}{n} \, , \qquad D_y = \sum_{i=1}^n \frac{D_{iy}}{n} \, ,
\ee
the car moving along the direction joining this point with the center of mass of the water, with initial velocity given by Eq.$\,$(\ref{v0lr}) and net force in (\ref{forcelr}) with $D=\sqrt{D_x^2 + D_y^2}$.
\end{quote}
In a sense, this means that the most general case can always be reduced to that in Fig.$\,$\ref{fig5}c. From a physical point of view, this theorem immediately follows from the fact the system never rotates by itself, along with the assumption that an equal mass of water (accounting to $M_0/n$) flows from each hole.

The situation of course changes when differently sized holes are present, but a generalization of the above theorem can be formulated.

For definiteness, let us consider a leaky tank car with two holes at distances $D_1$ and $D_2$ from the center and with surface area $S_1$ and $S_2$, respectively (1D restriction is enough). In the time interval $\Delta t$, the water flowing from the two holes is given by $\Delta V_{1,2}/\Delta t = S_{1,2} \, v_{E1,2}$. Since (within the usual assumptions) the pressure on the water exiting from the two holes is equal to that on the free upper surface (assumed to be equal to the atmospheric pressure), the efflux velocities $v_{E1,2}$ from the holes are as well equal, depending only on the height $h$ of the water inside the car:  $v_{E1}=  v_{E2}$. As a result,
\be
\frac{1}{S_1} \, \frac{\Delta V_1}{\Delta t} = \frac{1}{S_2} \, \frac{\Delta V_2}{\Delta t} \, ,
\ee
and since the conservation law for the volumetric flow requires that
\be
\frac{\Delta V_1}{\Delta t} + \frac{\Delta V_2}{\Delta t} = \frac{\Delta V}{\Delta t}
\ee
($\Delta V$ being the total volume flowing in the time $\Delta t$ from the car), we finally get:
\be
\frac{\Delta V_1}{\Delta t} = \frac{S_1}{S_1 + S_2} \, \frac{\Delta V}{\Delta t} \, , \qquad 
\frac{\Delta V_2}{\Delta t} = \frac{S_2}{S_1 + S_2} \, \frac{\Delta V}{\Delta t} \, .
\ee
By introducing the density $\rho$, we can easily translate such relations into analogous ones for the rate of water mass change from the given holes ($\dot{M}_{1,2} = \rho \Delta V_{1,2} / \Delta t$), that is $\dot{M}_{1,2} = S_{1,2} \, \dot{M} / (S_1+S_2)$ ($\dot{M}$ being the rate of mass change in the car). As discussed above, the momenta of the water acting on the car, and flowing from the two holes, are then given by
\be
p_{w_{\rm ins}1,2} = - \dot{M}_{1,2} \, \frac{D_{1,2}}{2} = - \, \frac{S_{1,2}}{S_1 + S_2} \, \dot{M} \, \frac{D_{1,2}}{2} \, ,
\ee
so that the constant forces on the tank car ``produced" by the two different holes are thus:
\be
F_1 = - \, \frac{S_1}{S_1 + S_2} \, \frac{\dot{M}_0 \, D_1}{2 \, t_s} \, , \qquad
F_2 = - \, \frac{S_2}{S_1 + S_2} \, \frac{\dot{M}_0 \, D_2}{2 \, t_s} \, .
\ee
Without dwelling in further mathematical details, we can then formulate the generalized center of mass theorem, whose formal proof is left to the interested student:
\begin{quote}
A leaky tank car problem with $n$ holes of different surface areas $S_1, S_2, \dots , S_n$, located at positions $\vec{D}_1, \vec{D}_1, \dots , \vec{D}_n$, is equivalent to a problem with just one hole located at
\be
\vec{D} = \frac{S_1 \vec{D}_1 + S_2 \vec{D}_2 + \dots S_n \vec{D}_n}{S_1 + S_2 + \dots S_n} 
\ee
with net force on the car given by
\be
\vec{F} = \frac{M_0 \, \vec{D}}{t_s^2} \, ,
\ee
and initial velocity
\be
\vec{v}_0 = \frac{\dot{M}_0 \, \vec{D}}{2(m + M_0)} \, .
\ee
\end{quote}

\section{Conclusions}

\noindent The leaky tank car problem is a unique opportunity to illustrate a number of physical issues concerning standard mechanics in a highly non-standard fashion, able to capture the interest of students involved in science at different levels. Here we have especially pointed out the intriguing physical aspects of the problem, although sometimes concealed behind mathematical properties,\cite{mcdonald, mcdonald2} by just employing only the strictly required mathematical tools, in order to be able to reach as much students as possible. In particular, we have shown that calculus is effectively required only to find the solution of the Newtonian equation of motion, or even to study the limiting cases of such solutions and related quantities, whereas only some (pseudo) numerical ability is claimed to build insightful plots. While undergraduate students are well able to take up this mathematical challenge (and we certainly urge them to do it), any student can on the other hand appreciate all the physical content of the problem. 

Indeed, in addition to the intriguing properties -- deduced within the simple model assumed and its generalization -- properly related to the motion of the leaky tank car (only already partly present in the literature), above we have as well presented the equally interesting features concerning the motion of the water flowing from the car, which certainly spurs physical reasoning in students even beyond the actual results obtained. Noteworthy and unexpected are a number of suitable generalizations of the problem, further stimulating students' interest, along with an important general theorem underlying the fact that water drainage never induces rotation. 

The approach followed here may be as well applied to other interesting problems, not limited to mechanics, with undoubted benefits for students and teachers.

\newpage

%\section*{Acknowledgments}
%\vspace{1truecm}

\appendix
\section{Analytic expansions in limiting cases}
\label{app1}

\noindent It could be a useful physically-inspired mathematical exercise for undergraduate students to obtain Taylor expansions of the solutions of the equation of motion in (\ref{aeq}-\ref{xeq}) in the different limiting cases considered in the present paper. Eqs.$\,$(\ref{aeq}-\ref{xeq}) are a good starting point for analytic expansions, provided that $t_s$ is kept finite, that is for finite $M_0$. Instead, while this is not true (for $M_0 \rightarrow 0, \infty$), the expressions mentioned should be re-written in terms of another finite time quantity (ruled by a finite $m$, rather than by $M_0$), which we define as 
\be
t_l \, = \, \sqrt{\frac{2 S_c m}{\rho S_h^2 g}} \, ,
\label{ttl}
\ee
according to which $t_s = \sqrt{\mu} \; t_l$. The resulting expressions replacing (\ref{aeq}-\ref{xeq}) are, then, the following:
\begin{align}
a(t) &= \frac{D}{t_l^2} \, \frac{1}{\dis 1 + \left( \sqrt{\mu} - \frac{t}{t_l} \right)^2} \, ,\label{aeql}
\\
v(t) &= - \, \frac{D}{t_l} \, \frac{\sqrt{\mu}}{1 + \mu} \left\{ 1 - \frac{1 + \mu}{\sqrt{\mu}} \left[ \arctan \sqrt{\mu} \vphantom{\left( 1 - \frac{t}{t_s} \right)} \right. \right. \nonumber \\ &  \quad \left. \left. - \arctan \left( \sqrt{\mu} - \frac{t}{t_l} \right) \right] \right\} , \label{veql} 
\\
x(t) &= - \, D \left\{ \frac{\sqrt{\mu}}{1 + \mu} \, \frac{t}{t_l} - \frac{1}{2} \left[ \log \frac{1 + \mu}{\dis 1 + \left( \sqrt{\mu} - \frac{t}{t_l} \right)^2} \right. \right. \nonumber \\ &  \quad - 2 \left( \sqrt{\mu} - \frac{t}{t_l} \right) \left( \arctan \sqrt{\mu} \vphantom{\left( 1 - \frac{t}{t_s} \right)} \right. \nonumber \\ &  \qquad \left.  \left. \left. - \arctan \left( \sqrt{\mu} - \frac{t}{t_l} \right) \right) \right] \right\} . \label{xeql} 
\end{align}

\begin{figure}
\bt{c}
\includegraphics[width=8.5cm]{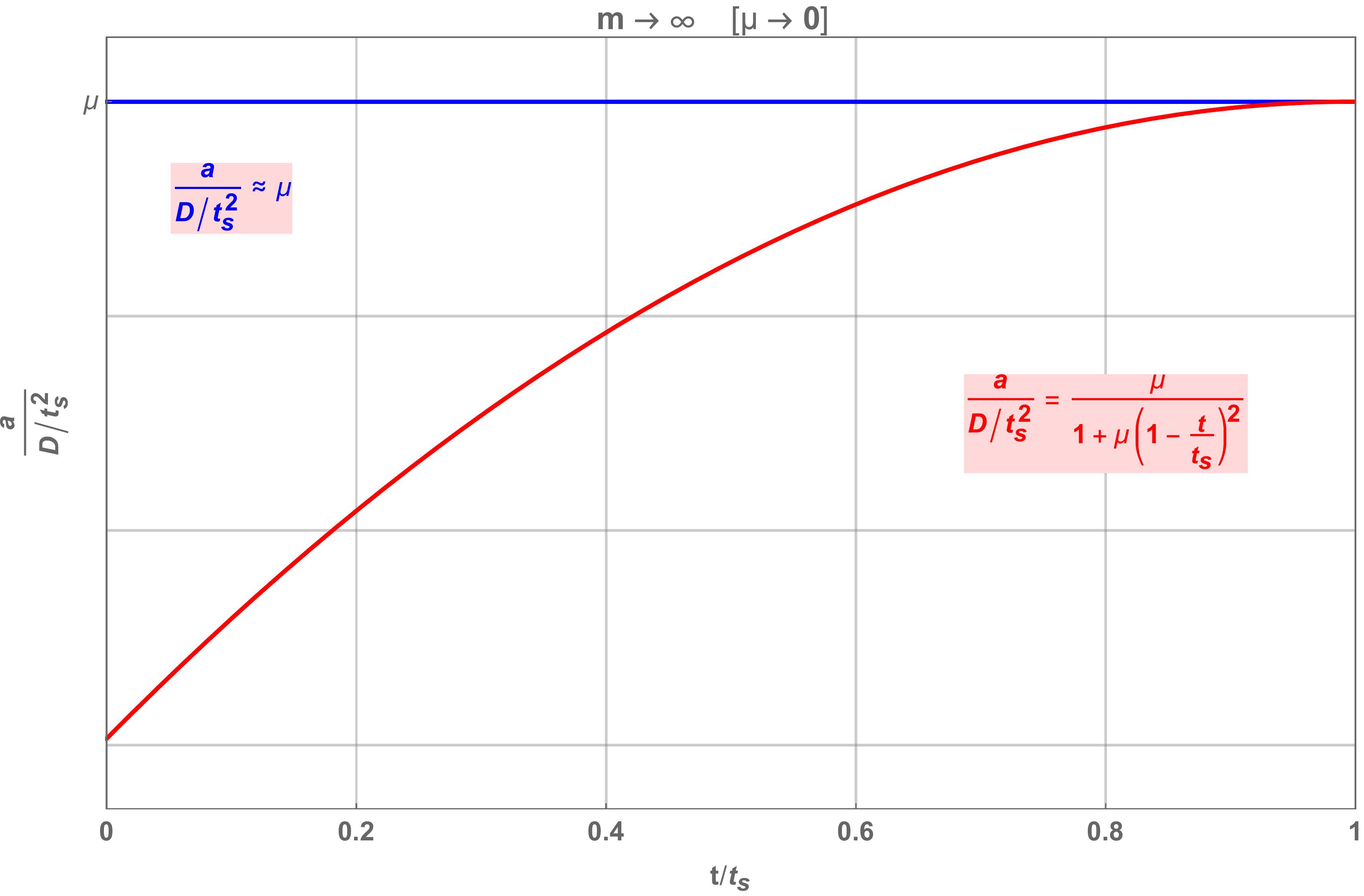} \\
\includegraphics[width=8.5cm]{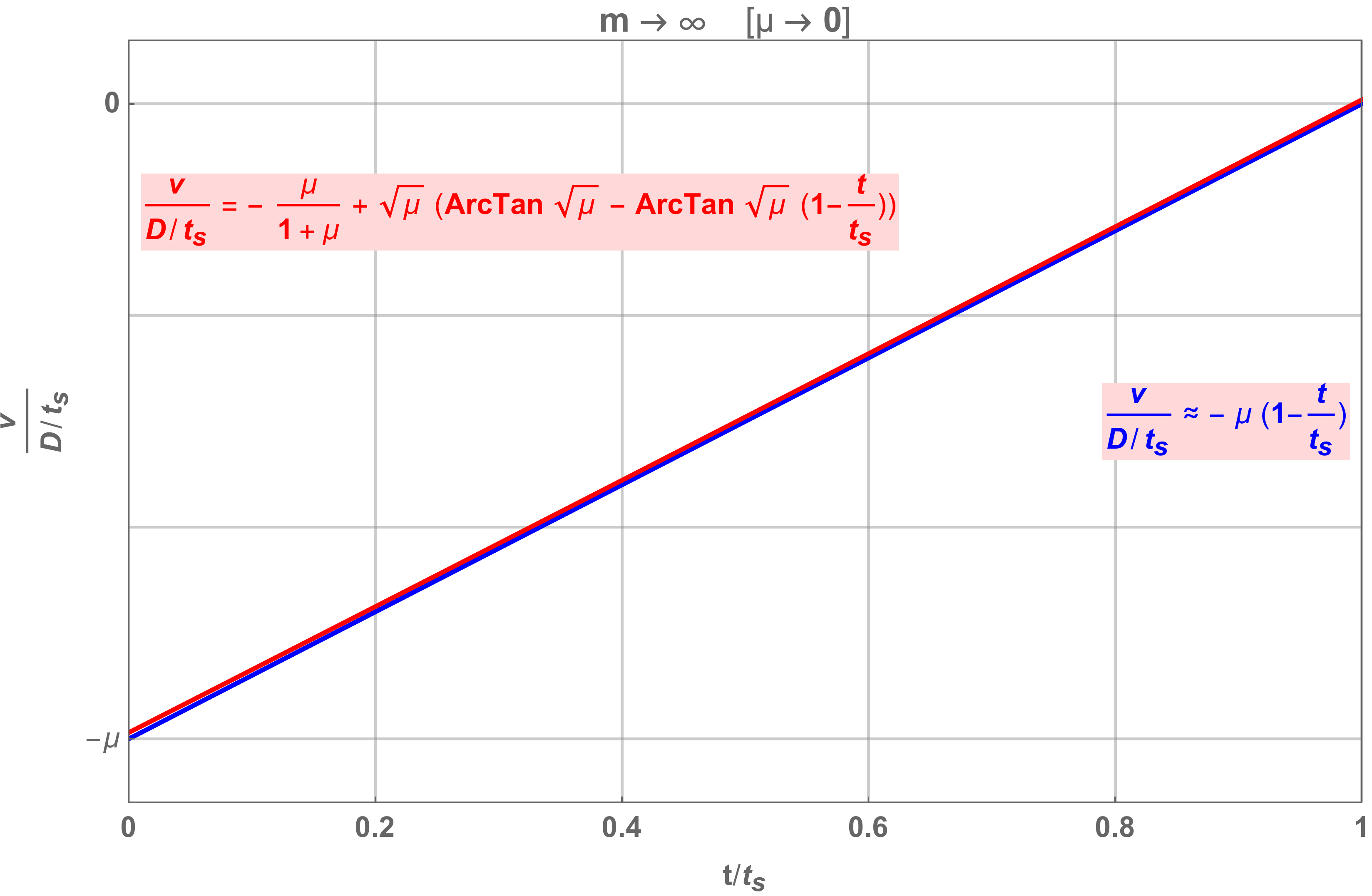} \\
\includegraphics[width=8.5cm]{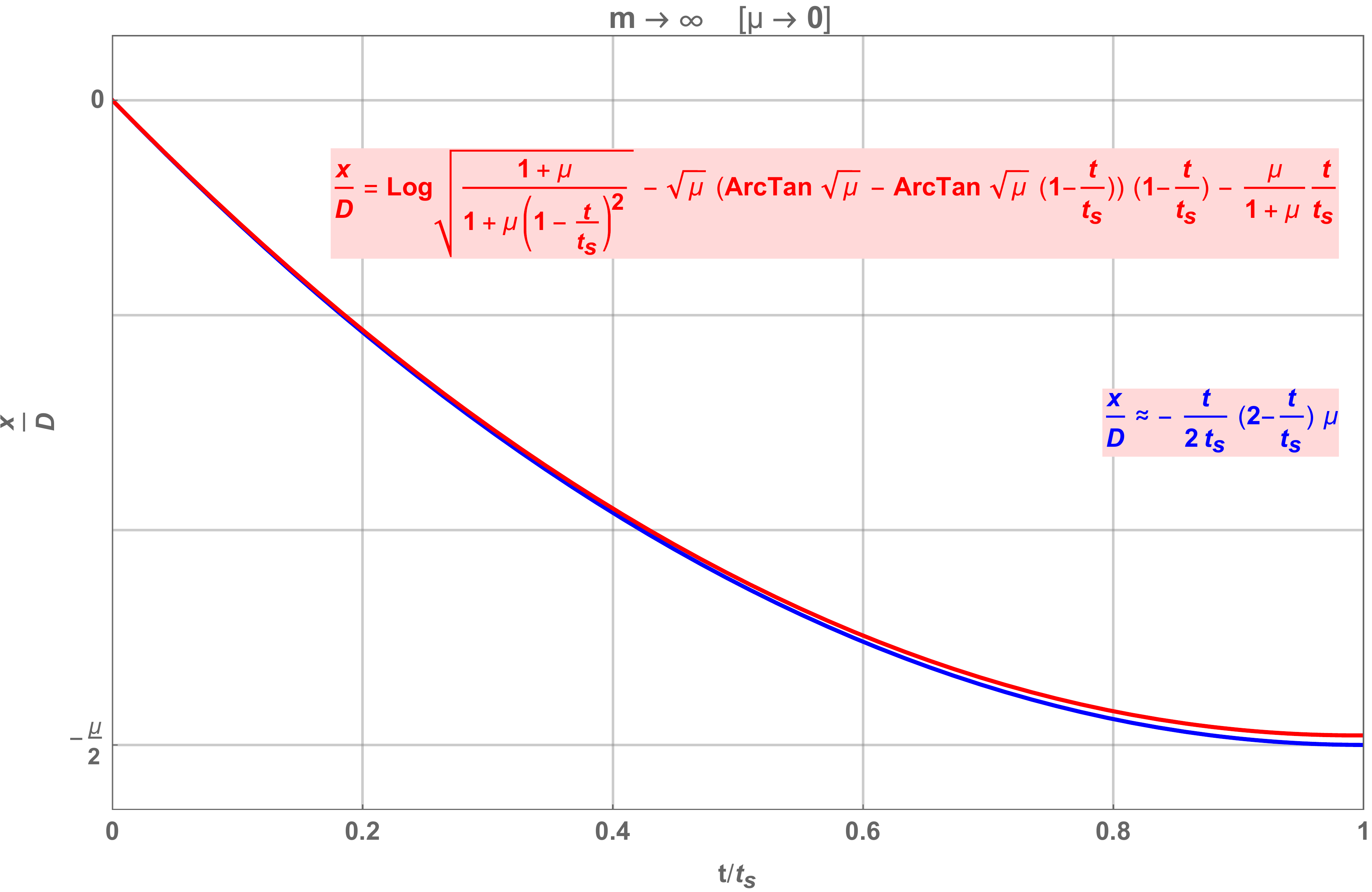} 
\et
\caption{Acceleration $a$, velocity $v$ and position $x$ of a leaky tank car as function of time, for large car masses ($m \rightarrow \infty$; $\mu \rightarrow 0$) and finite water mass ($M_0 \neq 0$). Red: exact expressions. Blue: approximate expansions at leading orders. In a finite emptying time ($t_s \neq 0$), for $\mu \rightarrow 0$ acceleration and velocity are negligible, the car covering just a small distance during its motion, without inversion.}
\label{figa1}
\end{figure}

\begin{figure}
\bt{l}
\includegraphics[width=7.4cm]{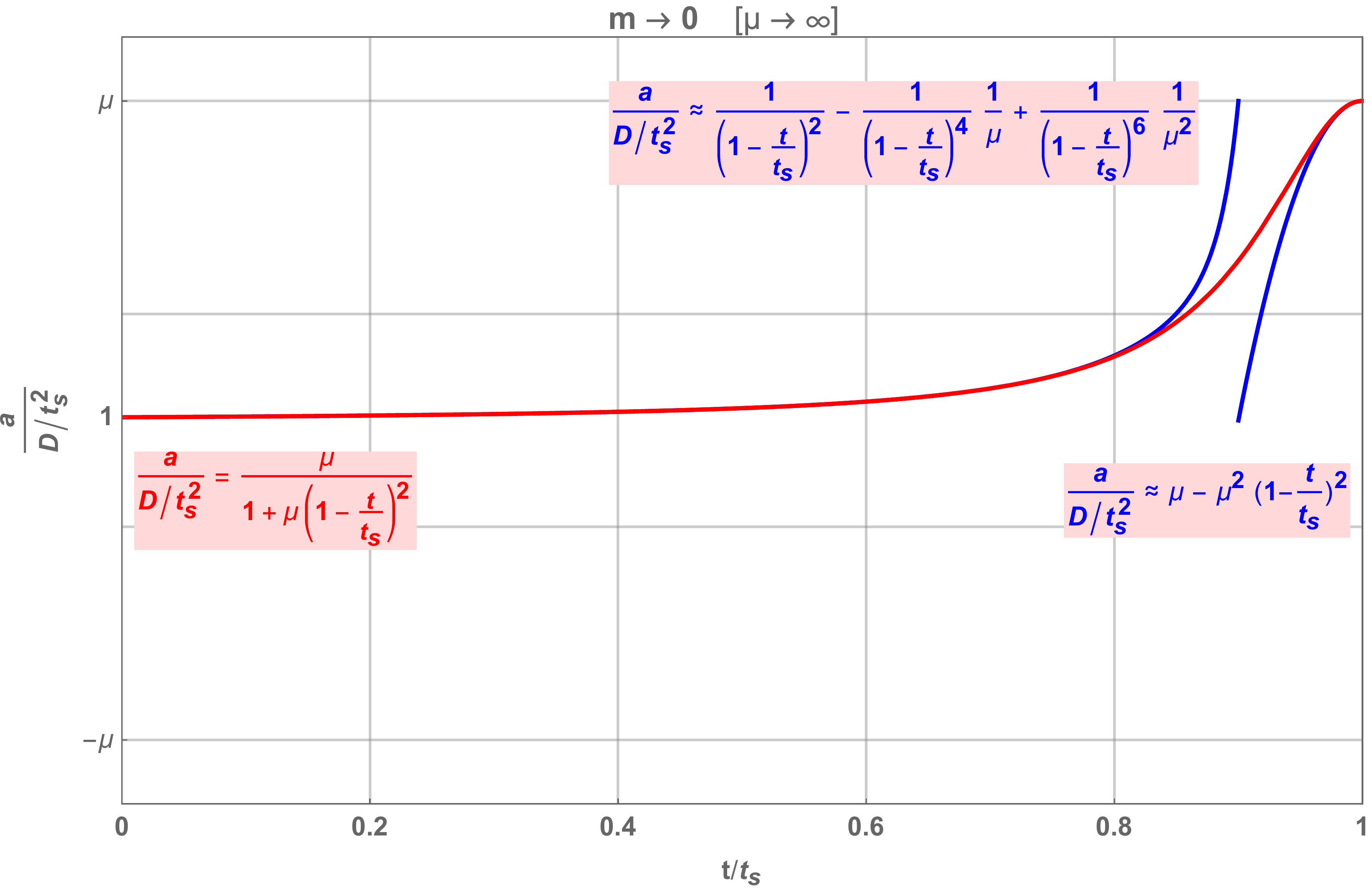} \\
\includegraphics[width=8.55cm]{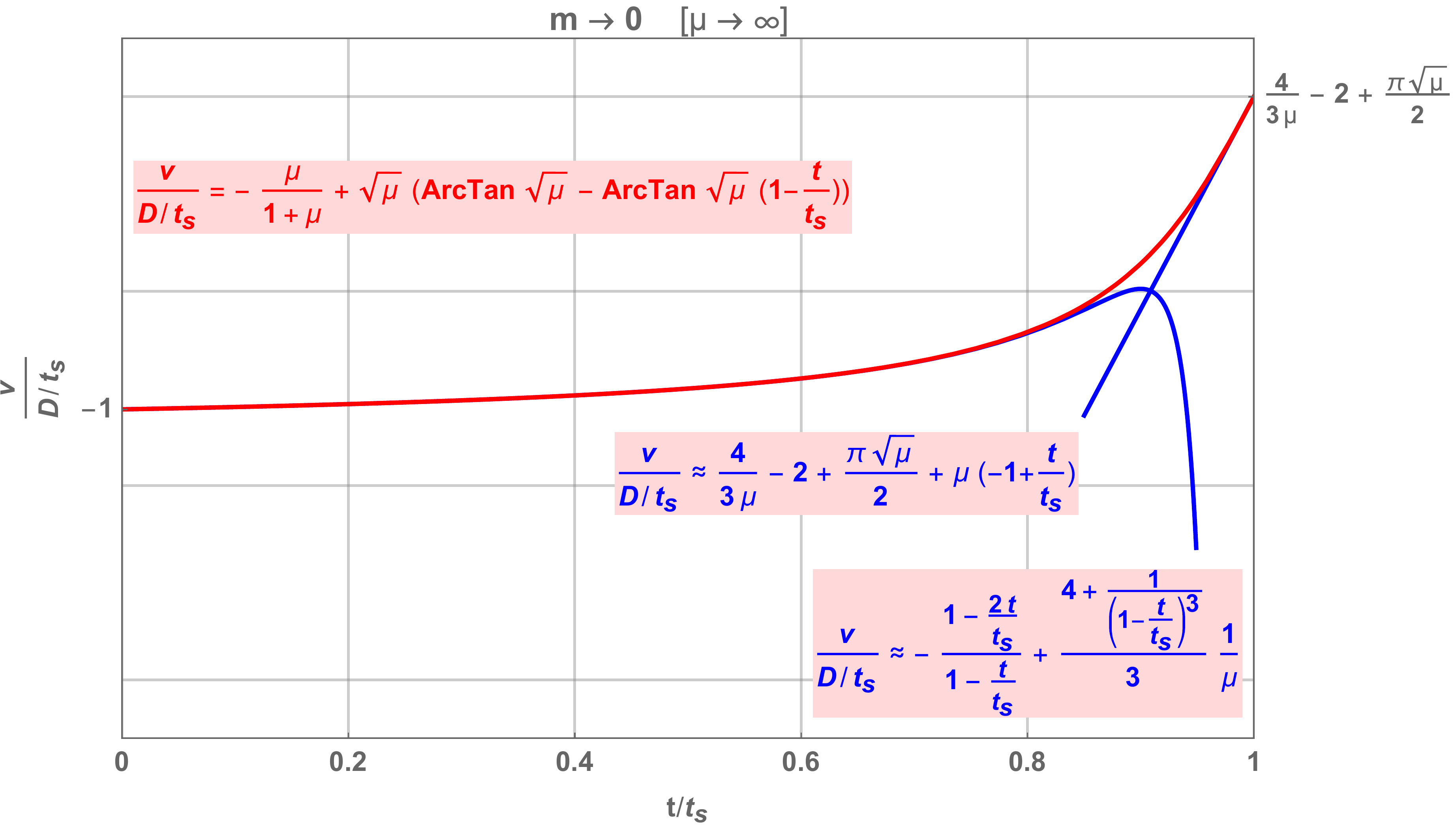} \\
\includegraphics[width=8.6cm]{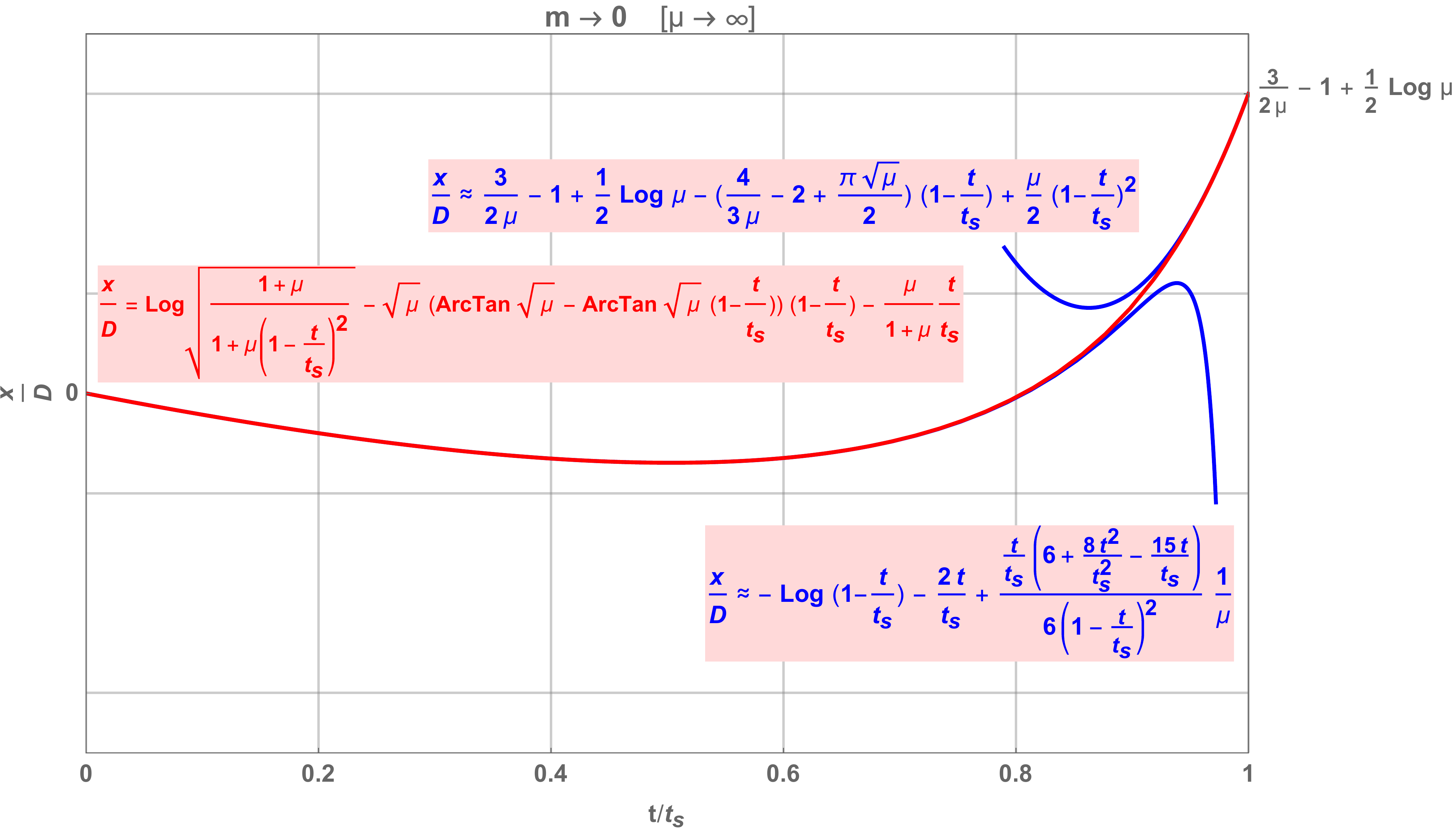} 
\et
\caption{Acceleration $a$, velocity $v$ and position $x$ of a leaky tank car as function of time, small car masses ($m \rightarrow 0$; $\mu \rightarrow \infty$) and finite water mass ($M_0 \neq 0$). Red: exact expressions. Blue: approximate expansions at leading orders. For $t \ll t_s$ and $t \approx t_s$, different expansions are required due to the cancelling of infinities for $t \rightarrow t_s$ and $\mu \rightarrow \infty$. The inversion of motion occurs at a time approximately equal to half of the (finite) emptying time $t_s$; before such value, the acceleration is roughly constant, while largely increasing (for $\mu \rightarrow \infty$) when $t_s$ is approaching. In such a case, the car covers a very large distance, ending with an increasingly large final velocity.}
\label{figa2}
\end{figure}

\begin{figure}
\bt{c}
\includegraphics[width=8.5cm]{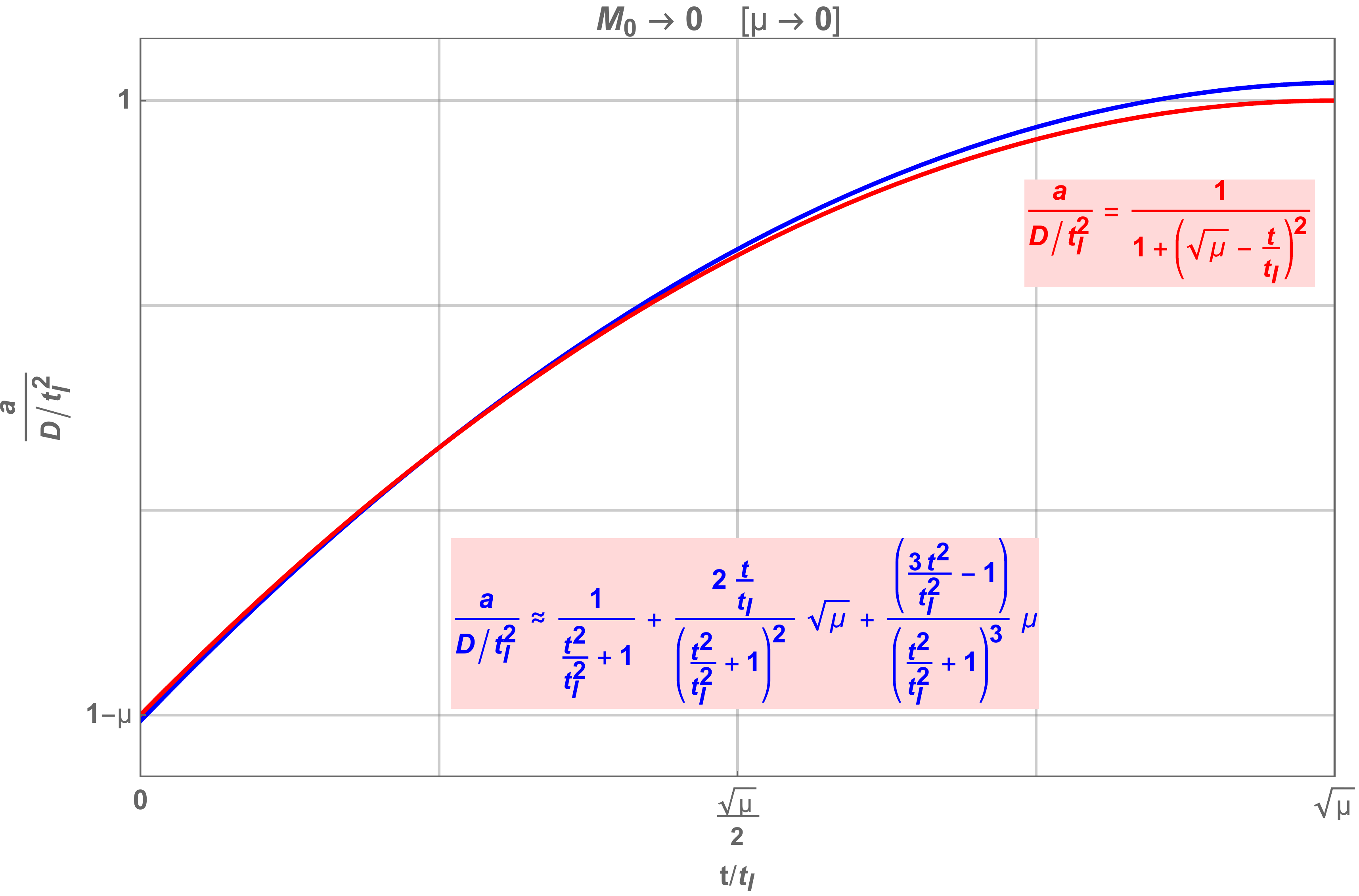} \\
\includegraphics[width=8.5cm]{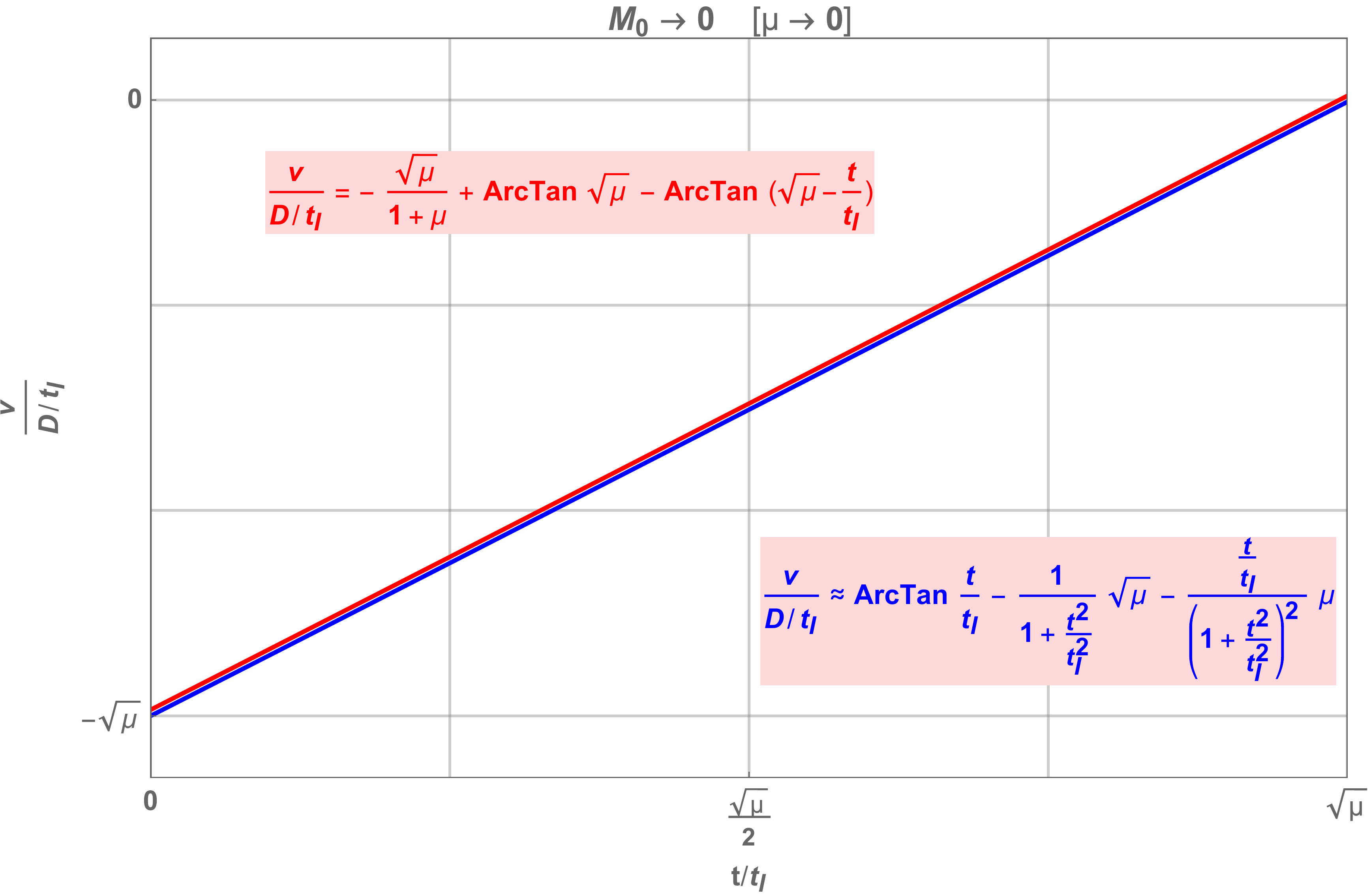} \\
\includegraphics[width=8.5cm]{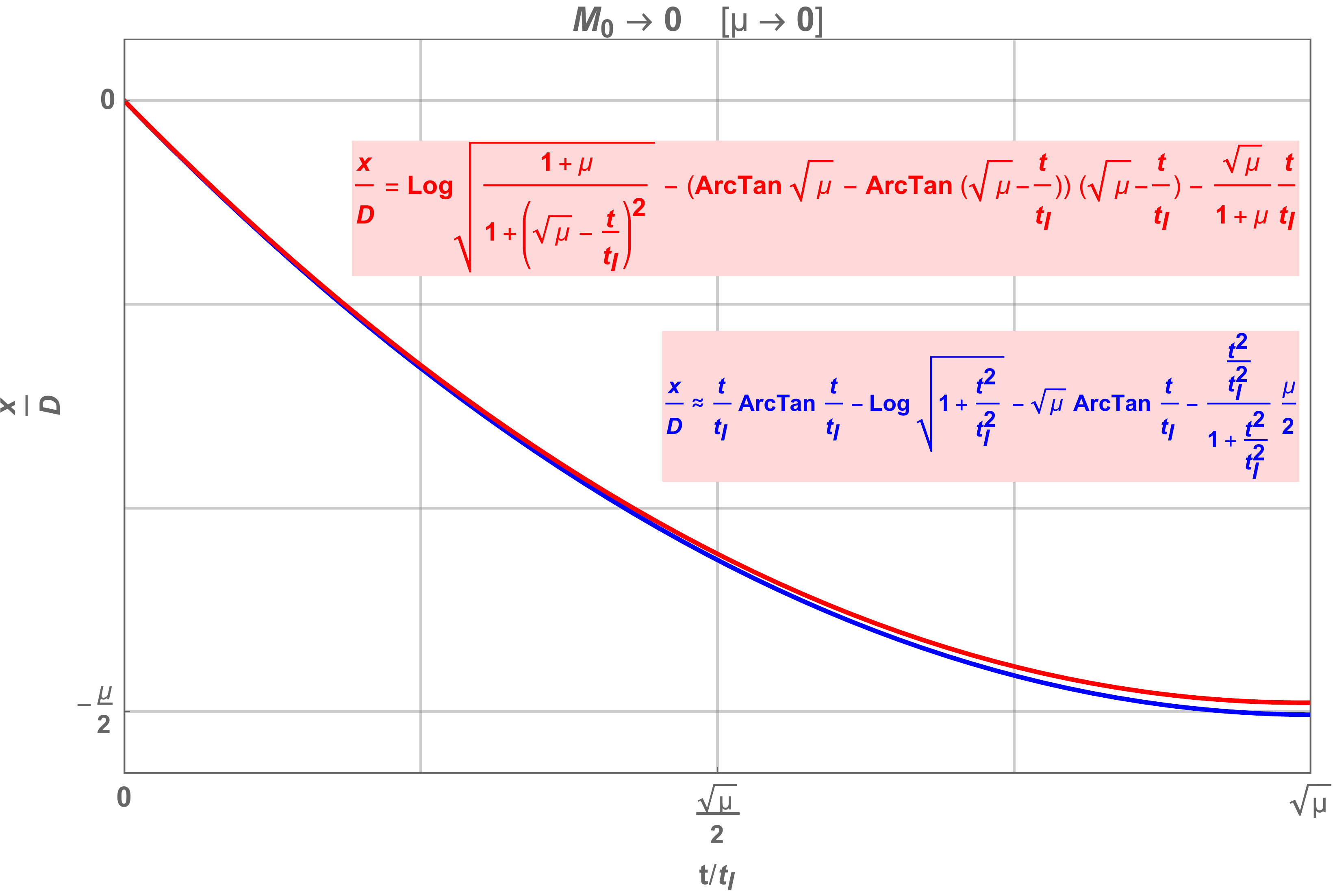} 
\et
\caption{Acceleration $a$, velocity $v$ and position $x$ of a leaky tank car as function of time, for small water masses ($M_0 \rightarrow 0$; $\mu \rightarrow 0$) and finite car mass ($m \neq 0$). Red: exact expressions. Blue: approximate expansions at leading orders. For $\mu \rightarrow 0$, acceleration is approximately constant but the velocity is negligibly small, the car covering just a small distance during its motion, without inversion, in the negligibly small emptying time ($t_s = \sqrt{\mu} \; t_l \rightarrow 0$).}
\label{figa3}
\end{figure}

\begin{figure}
\bt{c}
\includegraphics[width=8.5cm]{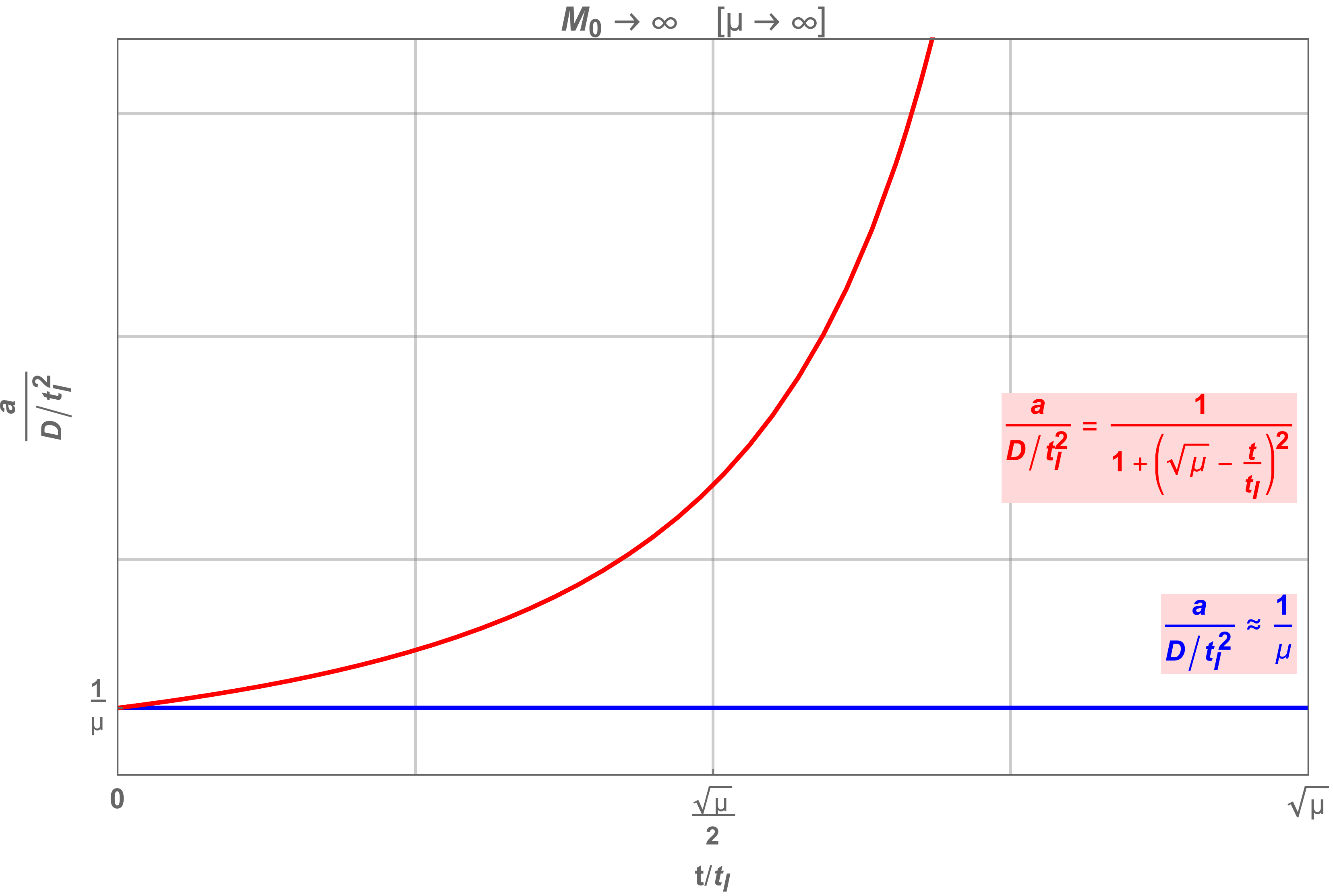} \\
\includegraphics[width=8.5cm]{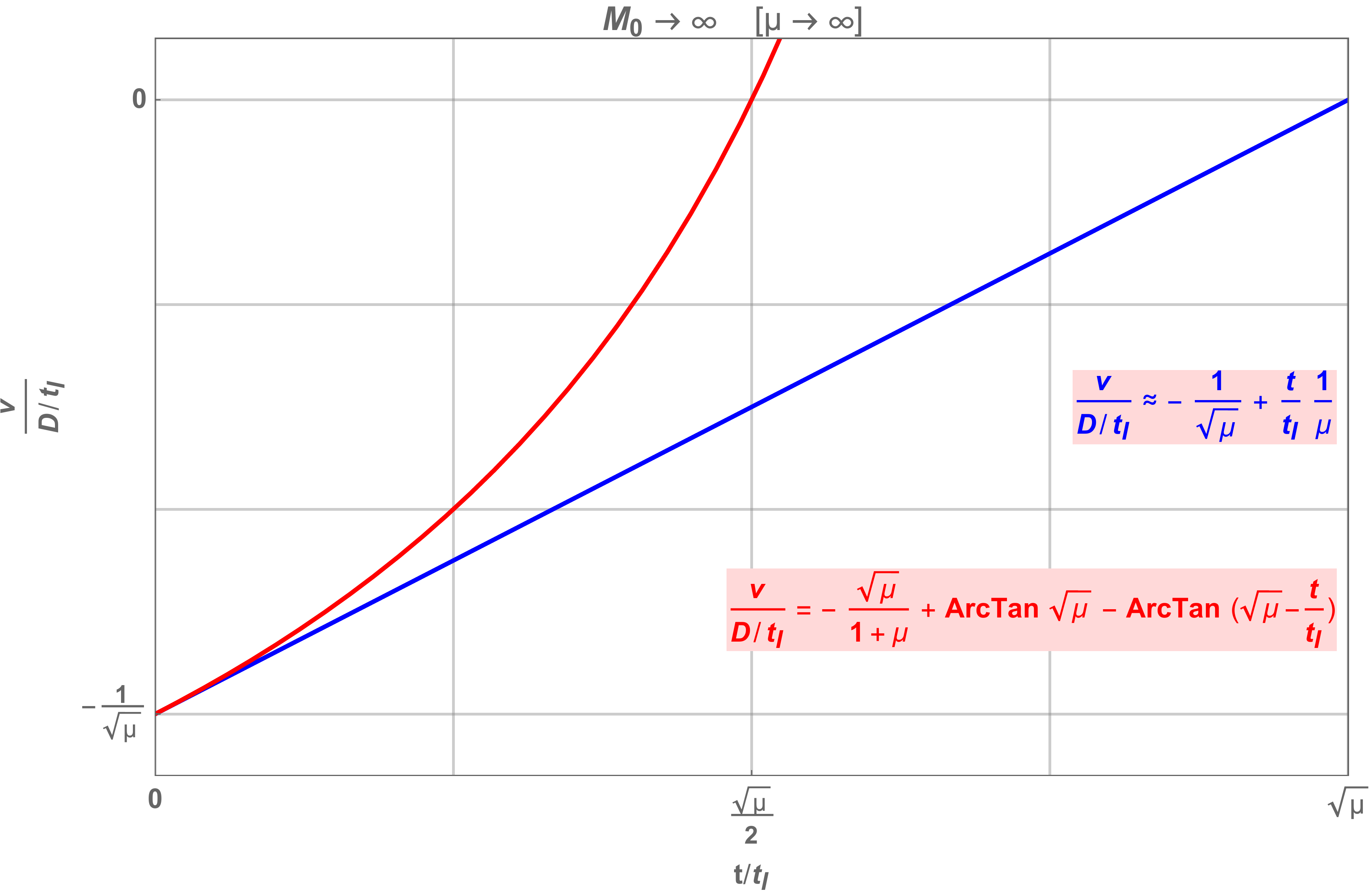} \\
\includegraphics[width=8.5cm]{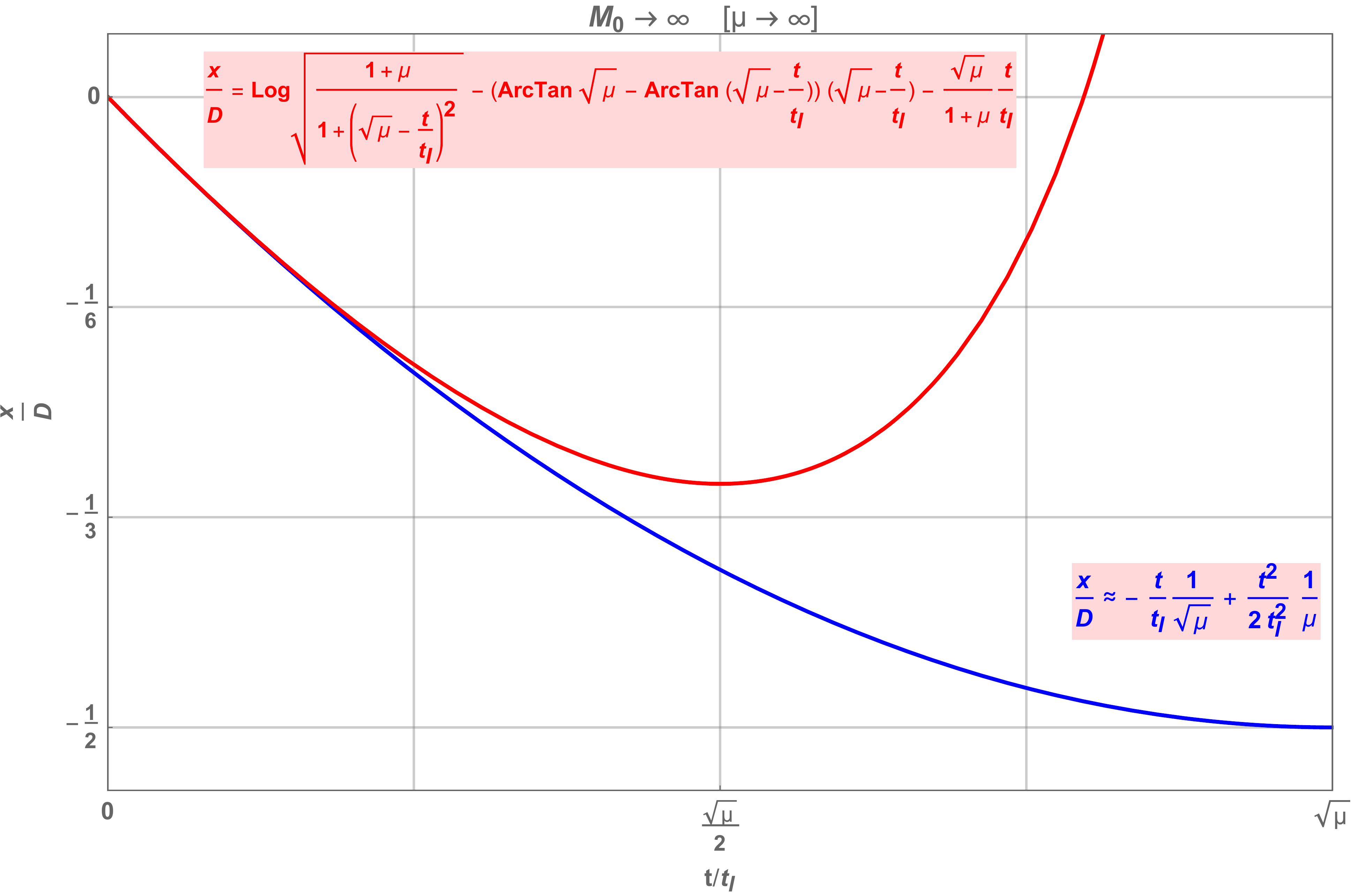} 
\et
\caption{Acceleration $a$, velocity $v$ and position $x$ of a leaky tank car as function of time, for large water masses ($M_0 \rightarrow \infty$; $\mu \rightarrow \infty$) and finite car mass ($m \neq 0$). Red: exact expressions. Blue: approximate expansions at leading orders. The emptying process takes an exceedingly large time ($t_s = \sqrt{\mu} \; t_l \rightarrow \infty$), during which practically no inversion can occur. The motion can be approximated by a uniformly accelerated one, although acceleration and velocity practically vanish in the limit $\mu \rightarrow \infty$, the car effectively covering a negligibly small distance.}
\label{figa4}
\end{figure}

We report in Fig.s \ref{figa1}-\ref{figa4} the results for the relevant expansions along with a plot of them, by highlighting in red color the exact expressions, while in blue color the approximate ones. It should be remarked that, in some cases (see figures), different expansions should be considered for a given case for $t \approx t_s$, due to potential cancellation of infinities occurring for such values in Eqs.$\,$(\ref{aeq}-\ref{xeq}).

It is as well a useful exercise for interested students to obtain these plots by themselves, by using some known computer program, or even to write an appropriate one for the different cases. Interpretation of the results, as highlighted in the main text, should then follow.

\section{Momentum contributions}
\label{app2}

\begin{figure*}[t]
\bt{ccc}
\includegraphics[width=8.4cm]{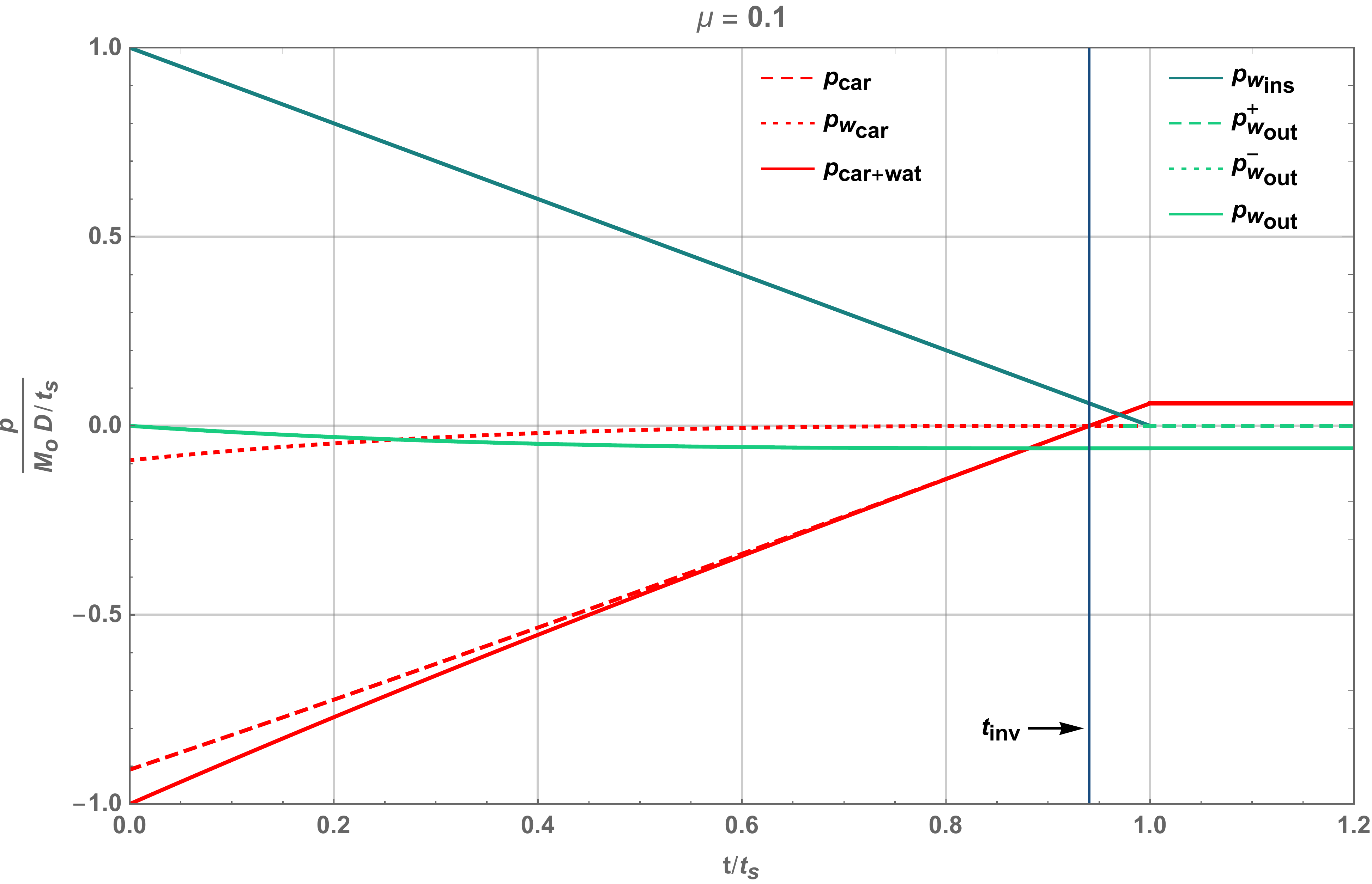} & $\qquad$ &
\includegraphics[width=8.4cm]{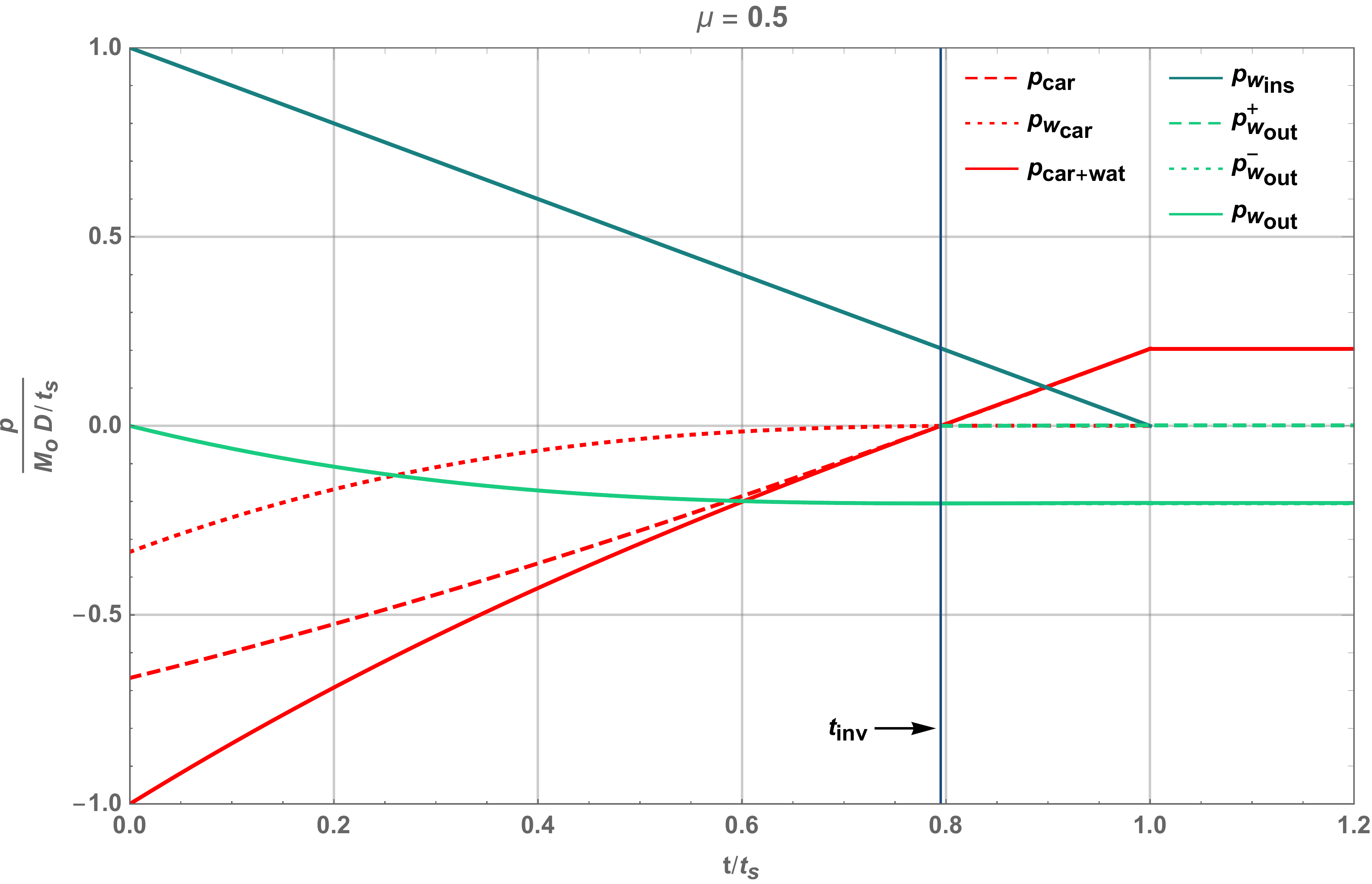} \\ & & \\
\includegraphics[width=8.4cm]{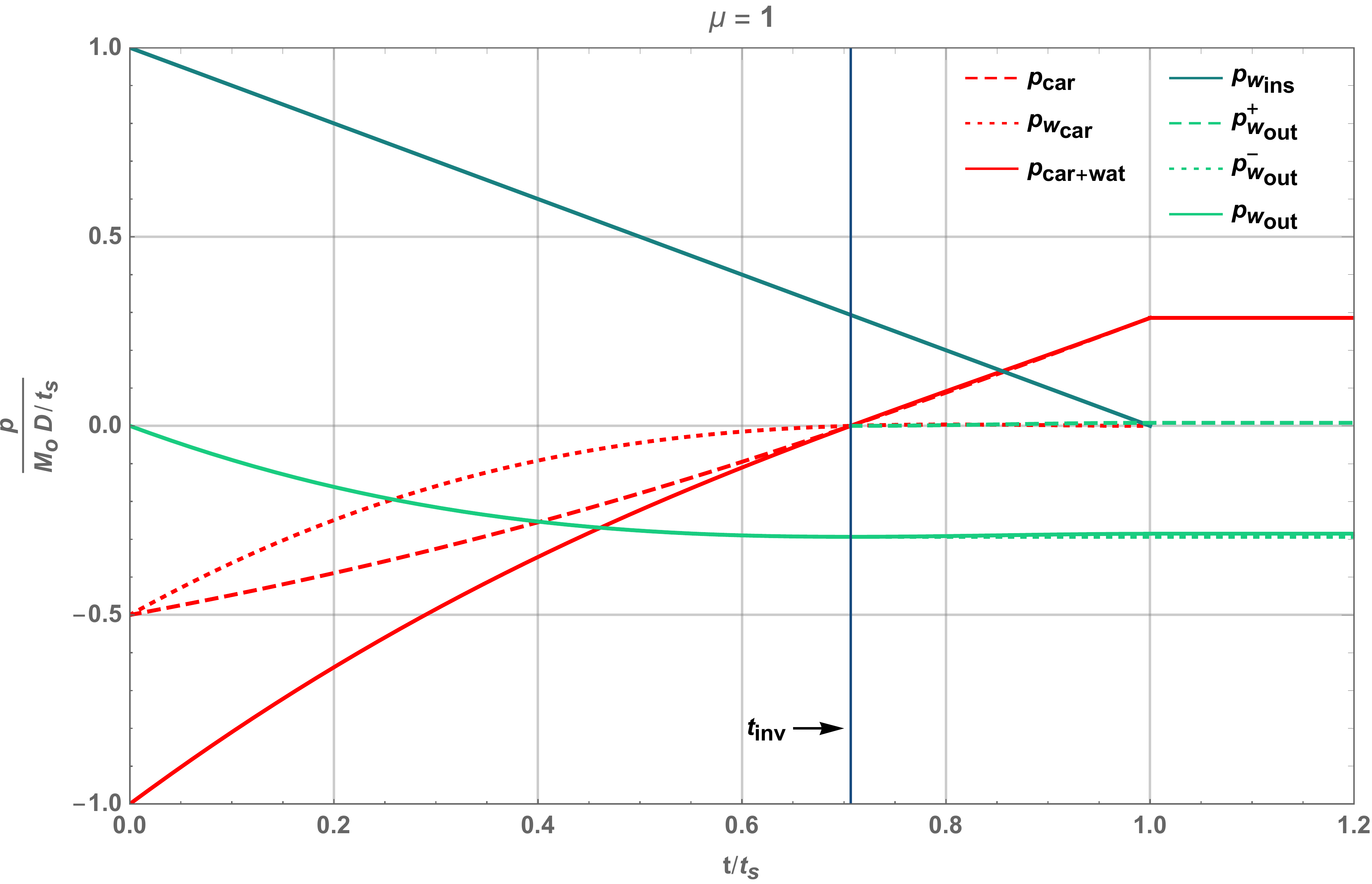} & $\qquad$ &
\includegraphics[width=8.4cm]{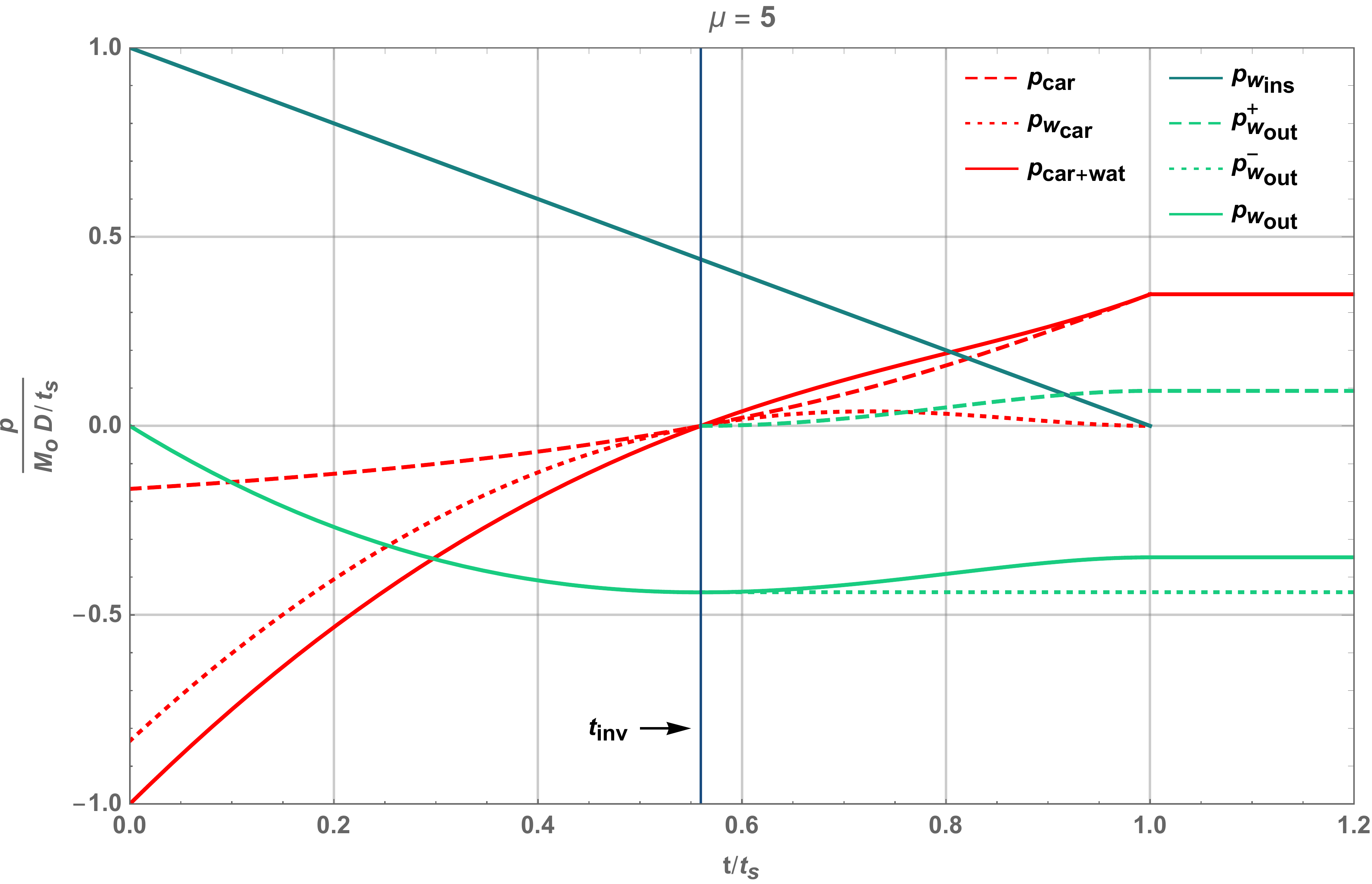} \\ & & \\
\includegraphics[width=8.4cm]{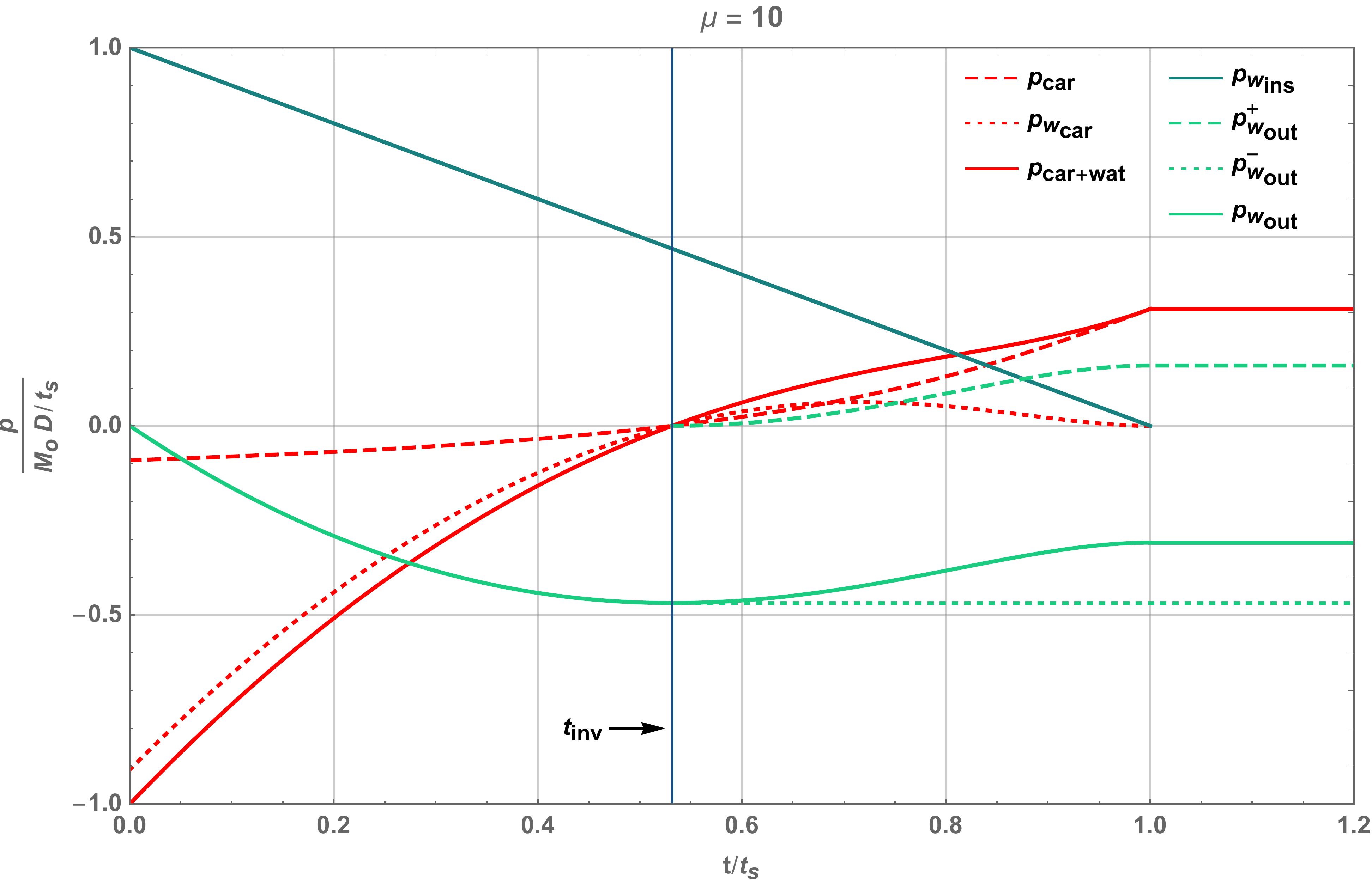} & $\qquad$ &
\includegraphics[width=8.4cm]{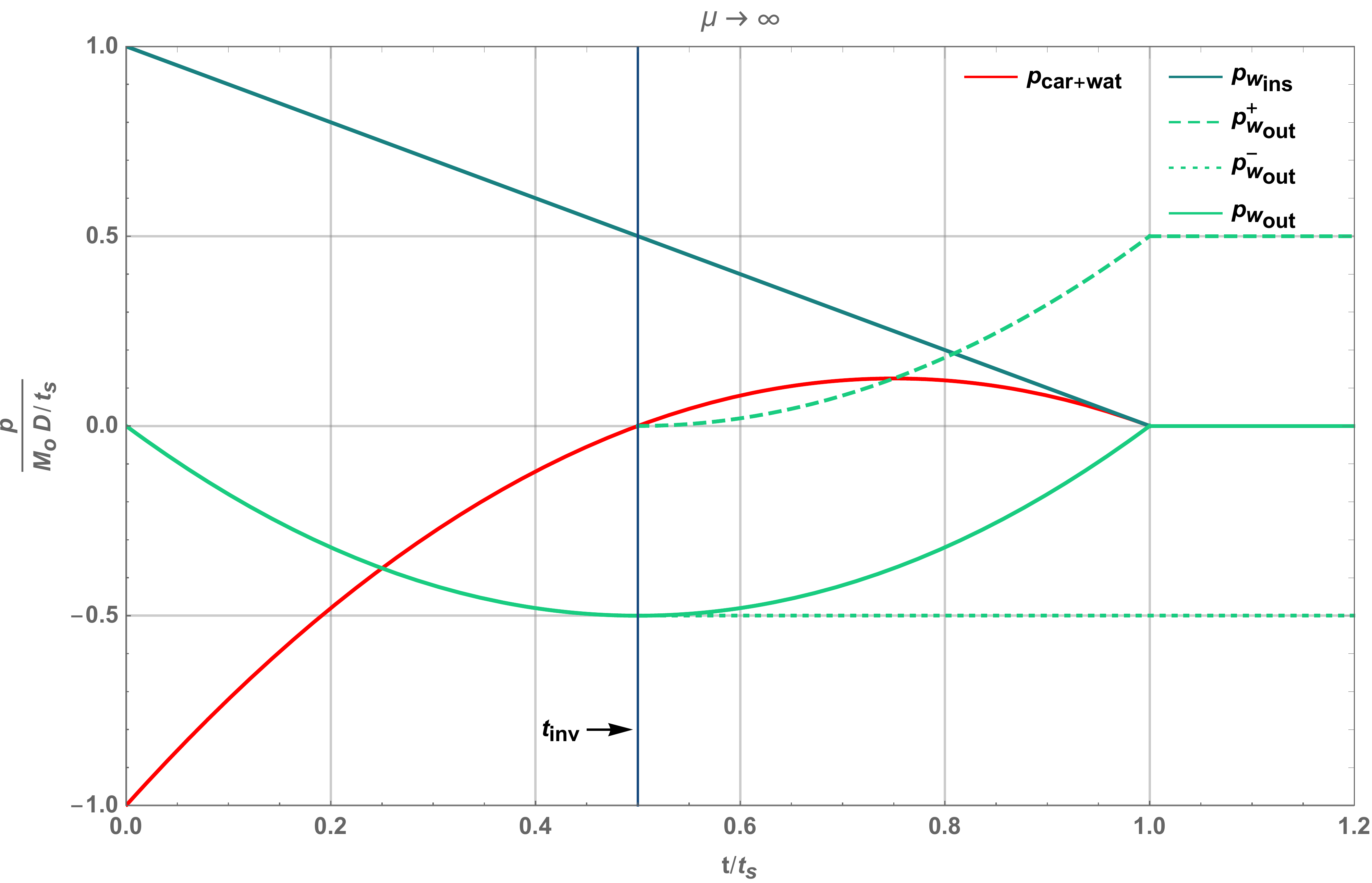}
\et
\caption{Momentum contributions (see text) to the leaky tank car problem as function of time, for different values of the mass ratio parameter $\mu$ (here, $p_{\rm car + wat} = p_{\rm car} + p_{w_{\rm car}}$). The time $t_{\rm inv}$ at which the inversion of the motion occurs is pointed out (for increasing $\mu$ it tends to the value $t_s/2$). For illustrative purposes, we also depict what happens for $t > t_s$, where the motion of the (empty) car and (leaked) water proceeds by inertia. The last figure, for $\mu \rightarrow \infty$, corresponds properly to the limit $m \rightarrow 0$, which is the only relevant limiting case (see discussion in the text).}
\label{figa5}
\end{figure*}

\noindent The leaky tank car problem accounts for four distinct contributions entering in the momentum conservation law, as discussed in Sect. \ref{initcond}. Once known the complete solution (\ref{veq}) of the equation of motion, the specific different weights of such contributions can be thoroughly analyzed, as done below. By introducing the auxiliary function:
\be
\varphi (t, \mu) = 1 - \frac{1 + \mu}{\sqrt{\mu}} \left[ \arctan \sqrt{\mu} - \arctan \sqrt{\mu} \left( 1 - \frac{t}{t_s} \right) \right] ,
\ee
from Eqs.$\,$(\ref{veq}), (\ref{mt}), (\ref{mpt}) we explicitly have:
\begin{align}
\frac{p_{\rm car}}{M_0 D/t_s} &= - \frac{1}{1 + \mu} \, \varphi(t, \mu) \, , \label{pcar} \\
\frac{p_{w_{\rm car}}}{M_0 D/t_s} &= - \frac{\mu}{1 + \mu} \left( 1 - \frac{t}{t_s} \right)^2 \varphi(t, \mu) \, , \label{pwcar} \\
\frac{p_{w_{\rm ins}}}{M_0 D/t_s} &= 1 - \frac{t}{t_s} \, , \label{pwins} \\
\frac{p_{w_{\rm out}}}{M_0 D/t_s} &= - \, \frac{2 \mu}{1 + \mu} \, \int_0^{t/t_s} \!\!\!\!\!\!\!\! (1 - z) \, \varphi (z t_s, \mu) \, \drm z \, . \label{pwout}
\end{align}
As explained in the paper, when the inversion of the motion occurs, it is interesting to consider separately the contributions to $p_{w_{\rm out}}$ due to the water exiting from the car and flowing (in the reference frame of Fig.$\,$\ref{fig2}) leftward ($-$) and rightward ($+$), given by:
\begin{align}
p_{w_{\rm out}}^-(t) &= \left\{ \ba{ll} p_{w_{\rm out}}(t) \, , & \quad 0 \leq t \leq t_{inv} \, , \\
p_{w_{\rm out}}(t_{\rm inv}) \, , & \quad t >	 t_{\rm inv} \, , \ea \right. \label{pwout-} \\
p_{w_{\rm out}}^+(t) &= \left\{ \ba{ll} 0 \, , & \quad 0 < t \leq t_{inv} \, , \\ p_{w_{\rm out}}(t) - p_{w_{\rm out}}(t_{\rm inv}) \, , & \quad t_{\rm inv} \leq t \leq t_s \, , \\
p_{w_{\rm out}}(t_s) -  p_{w_{\rm out}}(t_{\rm inv}) \, , & \quad t > t_s \, , \ea \right. \label{pwout+}
\end{align}
respectively. In Fig.$\,$\ref{figa5} we report all these contributions as function of time, for different values of the mass ratio parameter $\mu$, provided that inversion occurs. 

Here, as above, it would be a useful mathematical and numerical exercise for interested students to perform the integral in Eq.$\,$(\ref{pwout}), as well to obtain the relevant plots in Fig.$\,$\ref{figa5}. It is also particularly interesting to study all the limiting cases for small/large car/water masses, then interpreting the results obtained (compared with the last plot in Fig.$\,$\ref{figa5}).

\clearpage


\clearpage



\clearpage

\begin{thebibliography}{99}
%\doublespacing

\bibitem{ekman}
R. Ekman, ``Two toy models for the motion of a leaky tank car,'' arXiv:1906.04731 [physics.class-ph].

\bibitem{mcdonald}
K.T. McDonald, ``Motion of a leaky tank car,'' Am. J. Phys. {\bf
59}, 813-816 (1991).

\bibitem{mcdonald2}
The analysis by McDonald\cite{mcdonald} has been further enlarged by the author himself and his collaborators along the years (during which the leaky tank car problem has also been introduced in general physics courses), given its didactic relevance. See the update version at that author website: http://physics.princeton.edu/~mcdonald/examples/tankcar.pdf

\end{thebibliography}
\end{document}